\documentclass[12pt]{article}
\usepackage{amsmath,amssymb}
\usepackage{amsthm,xcolor}
\usepackage{times}
\usepackage{braket}
\usepackage{mathtools}
\usepackage{dsfont}
\usepackage{graphicx}
\usepackage{booktabs}
\usepackage{makecell}
\usepackage{subfig}
\usepackage{hyperref} 
\hypersetup{
    colorlinks=true,
    linkcolor=blue,
    urlcolor=blue,
    citecolor=blue,
}
\usepackage[capitalise]{cleveref}
\usepackage[numbers,sort&compress]{natbib}
\usepackage{etoolbox} 
\usepackage{changes} 
\usepackage[thicklines]{cancel}

\usepackage{soul} 

\usepackage{multirow}
\usepackage{adjustbox}

\definecolor{ForestGreen}{rgb}{0.13,0.55,0.13}
\newcommand{\email}[1]{\href{mailto:#1}{\texttt{\color{blue}#1}}}

\allowdisplaybreaks[2]
\makeatletter

\newcommand{\Rmnum}[1]{\expandafter\@slowromancap\romannumeral #1@}
\makeatother

\def\tr {\mathrm{Tr}}

\newtheorem{theorem}{Theorem}

\newtheorem{lemma}{Lemma}

\title{Witness High-Dimensional Quantum Steering via Majorization Lattice
}

\author
{Ma-Cheng Yang$^{1}$ and Cong-Feng Qiao$^{1,2}$\footnote{correspondence: \email{qiaocf@ucas.ac.cn}}\\
\\
\normalsize{$^{1}$School of Physical Sciences, University of Chinese Academy of Sciences}\\
\normalsize{1 Yanqihu East Rd, Beijing 101408, China}\\
\normalsize{$^{2}$ICTP-AP, University of Chinese Academy of Sciences, Beijing 100190, China}
}

\date{}

\begin{document}

\baselineskip24pt
\maketitle

\begin{abstract}
Quantum steering enables one party to influence another remote quantum state by local measurement. While steering is fundamental to many quantum information tasks, the existing detection methods in the literature are mainly constrained to either specific measurement scenario or low-dimensional systems. In this work, we propose a majorization lattice framework for steering detection, which is capable of exploring the steering in arbitrary dimension and measurement setting. Steering inequalities for two-qubit states, high-dimensional Werner states and isotropic states are obtained, which set even stringent bars than what has been reached yet. Notably, the known high-dimensional results turn out to be some kind of approximate limits of the new approach.
\end{abstract}

\section*{Introduction}\label{sec:intro}
\noindent
The concept of quantum steering can be traced back to the renowned debate on the completeness of quantum mechanics initiated by Einstein, Podolsky, and Rosen in 1935 \cite{einstein35}, along with the subsequent responses from Schrödinger \cite{schrodinger35}. Quantum steering refers to the ability of one party (Alice) in a composite quantum system to influence the quantum state of another party (Bob) through local measurements. This phenomenon implies a distinct form of non-local correlations that are found to exist between quantum entanglement and Bell non-locality \cite{wiseman07,quintino15,uola20}. 

Recently, a lot of efforts have been made to characterize and understand \deleted{the} quantum steering by means of local hidden state (LHS) models and steering inequalities. In seminal works \cite{wiseman07,jones07}, Wiseman \emph{et al.} established the steering thresholds for Werner states and isotropic states under projective measurements by constructing LHS models (c.f. also Ref. \cite{werner89,almeida07}). Subsequent advances by Nguyen and Gühne determined the new thresholds for the restricted case of dichotomic measurements performed by Alice \cite{nguyen20}. Very recently, people found the Wiseman \emph{et al.}'s threshold condition remains valid for positive operator-valued measurements (POVMs) in qubit systems \cite{zhang24,renner24}, which challenges the presumed advantage of POVMs in quantum steering scenario. Moreover, similar to Bell non-locality, quantum steering can be identified as well by the violation of some inequalities. This provides a practical way to detect the quantum steering in experiments. To this aim, various steering inequalities have been proposed, including the well-known Reid criteria \cite{reid89}, linear steering inequalities \cite{saunders10,zheng17}, and steering inequalities based on various uncertainty relations \cite{walborn11,schneeloch13,zhen16,maity17,riccardi18,costa18,krivachy18,li21-nonlocality}, among others (see reviews \cite{cavalcanti17-Steering,uola20,xiang22}). 
Due to the equivalence between quantum steering and incompatibility of measurements \cite{uola15}, the results of measurement incompatibility can also be employed to characterize quantum steering \cite{designolle19}. In recent years, high-dimensional quantum steering has garnered significant attention due to its enhanced robustness against noise and loss \cite{skrzypczyk15,skrzypczyk18,zeng18,srivastav22,qu22}, while limited inequalities are established, merely applicable to qubit system or systems with specific measurement settings, such as mutually unbiased bases (MUBs).

Over the past few decades, majorization theory has found widespread applications across various aspects of quantum information science \cite{nielsen99,nielsen01,nielsen01-gxIX4,guhne04,partovi12,partovi11,friedland13,puchala13,rudnicki14,li16,li19,puchala18}. Of particular significance is the majorization lattice \cite{cicalese02,alberti82,bapat91,bondar94}, which establishes an algebraic structure for the set of probability distributions, providing a framework that is naturally congruent with the probabilistic descriptions inherent to quantum theory. Recently, majorization theory has been applied to quantum steering \cite{li20,li21-nonlocality,zhu23} in lower-dimensional systems.

In this paper, we present a systematic approach to the detection of quantum steering based on the probability majorization lattice. Notably, this method is applicable to any measurement setting and any dimension, by which the quantum correlation information encoded in joint probabilities can then be fully extracted through tailored aggregation operations. This leads to lossless information extraction in contrast to conventional methods relying on functions of probability distributions, such as entropy and variance. To illustrate the efficacy of our approach, we derive steering inequalities for two-qubit systems, as well as for arbitrary-dimensional Werner states and isotropic states. In the MUBs scenario, our approach not only yields stronger steering inequalities compared to existing ones, but also indicates that previous high-dimensional steering inequalities can be systematically obtained in a series of approximations. In the following we present first the preliminary preparation for later use, then the derivation of our central theorem, and representative examples confronting to the yet known results.

\section*{Results}
\subsection*{Preliminaries}

Majorization theory establishes a partial ordering relation between any two vectors $\vec{x},\vec{y}\in \mathds{R}^n$ with $\sum_{i=1}^{n}x_{i} = \sum_{i=1}^{n}y_{i}$, denoted $\vec{x}\prec\vec{y}$, i.e. $\vec{x}$ is majorized by $\vec{y}$, if and only if \cite{marshall79}
\begin{align}
&\sum_{i=1}^{k}x_{i}^{\downarrow} \leq \sum_{i=1}^{k}y_{i}^{\downarrow} \; , \; k=1,\cdots,n-1 \; .
\label{eq:majorization_def}
\end{align}
Here, the superscript $\downarrow$ denotes the components of $\vec{x}$ and $\vec{y}$ in decreasing order. For $n$-dimensional probability vectors with components in decreasing order, we can define
\begin{align}
\mathcal{P}_{n} \equiv\left\{\left(p_{1},\cdots,p_{n}\right)^{\mathrm{T}}\Big|p_{i}\geq p_{i+1}\geq 0,\sum_{i=1}^{n}p_{i}=\text{const.}\right\} \; .
\end{align}
The set $\mathcal{P}_n$, equipped with the majorization relation defined in \cref{eq:majorization_def}, forms a lattice \cite{cicalese02} (see \cite{davey02} for \added{a} thorough discussion). In mathematics, a lattice is defined as a quadruple, and a \emph{probability majorization lattice} can be represented as $\braket{\mathcal{P}_n, \prec, \vee, \wedge}$, where for $\forall \vec{p},\vec{q}\in\mathcal{P}_n$ there is a unique infimum $\vec{p}\wedge\vec{q}$ and a unique supremum $\vec{p}\vee\vec{q}$. It is proved that the probability majorization lattice is a complete lattice and we have \cite{alberti82,bapat91,bondar94}
\begin{lemma}\label{lem:prob_maj_lattice}
A probability majorization lattice is a complete lattice, meaning that for every subset $S\subseteq\mathcal{P}_n$, both the supremum $\bigvee S\in\mathcal{P}_n$ and the infimum $\bigwedge S\in\mathcal{P}_n$ exist. 
\end{lemma}

Probability majorization lattice provides an elegant formalism to establish state-independent uncertainty relation (UR). For any observables $A,B$ in a $d$-dimensional Hilbert space $\mathcal{H}$, one can define the measurement distribution set as
\begin{align}
\mathcal{P}_{A,B} = \{\vec{p}_A\odot\vec{p}_B|\vec{p}_A\in \mathcal{P}_A,\vec{p}_B\in\mathcal{P}_B\} \; .
\end{align}
Here, $\mathcal{P}_A$ and $\mathcal{P}_B$ are respectively the sets of probability vectors corresponding to the measurements $A$ and $B$, such as $\mathcal{P}_{A} = \{\left(p(a_{1}|A),\cdots,p(a_{d}|A)\right)^{\mathrm{T}}|\rho\in \mathcal{D}(\mathcal{H})\}$ with the set of density matrices $\mathcal{D}(\mathcal{H})$ for $d$-dimensional Hilbert space $\mathcal{H}$; $\odot$ denotes any binary operation that preserve the majorization relation, including direct sum, direct product, and vector sum, i.e. $\odot\in\{\otimes,\oplus,+\}$. In light of \cref{lem:prob_maj_lattice}, a probability majorization lattice is a complete lattice and there exists a supremum for the subset $\mathcal{P}_{A,B}\subset\mathcal{P}_{n}$, which formulates the majorization-based state-independent UR:
\begin{align}
\forall \vec{\chi}\in \mathcal{P}_{A,B} \; , \; \vec{\chi}\prec \vec{\omega}(A,B) \; ,
\end{align}
where $\vec{\omega}(A,B)=\bigvee \mathcal{P}_{A,B}$. Clearly, the statements above can be generalized to the scenario involving 
$N$ measurements. Note that the bound $\vec{\omega}(A,B)$ with binary operations $\{\otimes,\oplus\}$ has been extensively studied in references \cite{partovi11,friedland13,puchala13,rudnicki14,li16,li19,puchala18,li20}. Let $I_\mu\subset [d]$ be a subset of $[d]$, where $[d]$ be the set of all integers from 1 to $d$. For binary operation $\oplus$, we have $\vec{\omega}(\mathcal{B})=(\Omega_{1},\Omega_2-\Omega_1,\cdots,\Omega_{Nd}-\Omega_{N(d-1)},0,\cdots,0)$, where $\Omega_{L}$ for $N$ measurements $\mathcal{B}=(B_1,\cdots,B_N)$ is defined as
\begin{align}
\Omega_{L}
&\equiv\max_{\substack{I_1,\cdots,I_N \\ \sum_{\mu=1}^{N}|I_\mu|=L}}\max_{\rho}\tr\left[\rho\left(\sum_{\mu=1}^L\sum_{i\in I_\mu}\ket{b_i^{(\mu)}}\bra{b_i^{(\mu)}}\right)\right] \\ 
&= \max_{\substack{I_1,\cdots,I_N \\ \sum_{\mu=1}^{N}|I_\mu|=L}}\left\|\sum_{\mu=1}^L\sum_{i\in I_\mu}\ket{b_i^{(\mu)}}\bra{b_i^{(\mu)}}\right\| \; .
\end{align}
Here, $\|\cdot\|$ denotes the operator norm and $\ket{b_i^{(\mu)}}\bra{b_i^{(\mu)}}$ is $\mu$-th projector of measurement base $B_\mu$ (see Methods for more details).


\subsection*{Probability majorization lattice and quantum steering}

Assuming that Alice and Bob each possess one of the two subsystems of a bipartite state and measure on bases $A$ and $B$ respectively, in projective measurement, quantum theory predicts the joint probability 
\begin{align}
p(a_i,b_j|A,B;\rho) = \tr[\Pi_{ij}\rho] \; 
\end{align}
with projectors $\Pi_{ij}= \ket{a_i}\bra{a_i}\otimes \ket{b_j}\bra{b_j}$. Considering of all alternative measurements, a bipartite quantum state corresponds to the infinite set of joint probability $p(a_i,b_j|A,B;\rho)$, i.e.
\begin{align}
\rho \to \mathcal{P} = \{p(a_i,b_j|A,B;\rho)|\forall A,B\} \; .
\end{align}
If the LHS description does not exist, i.e. $p(a_i,b_j|A,B;\rho) \neq \sum_{\xi}\wp(a_i|A,\xi)\tr[\Pi_{j}^{B}\rho_{\xi}^B]\wp_{\xi}$ for certain measurements $A$ and $B$, Alice can then demonstrably steer Bob’s state, hence exhibiting the quantum steering \cite{wiseman07}. Here, $\{\xi,\wp_{\xi}\}$, $\wp(a_i|A,\xi)$ and $\rho_{\xi}^B$ denote the possible hidden variables, Alice's measurement distribution and Bob's local state, respectively. 


In practical experiments, only a finite number of measurement settings can be implemented. We define the set of the joint distributions of non-steerable states in $N$-measurement scenario as follows:
\begin{align}
\mathcal{P}_{\mathrm{ns}}^{N} = \left\{\bigodot_{\mu=1}^N\vec{p}(A_\mu\otimes B_\mu)_{\rho} \bigg|\rho\in \mathcal{D}_\mathrm{ns}\right\} \; .
\end{align}
Here, $\mathcal{D}_\mathrm{ns}$ signifies the set of non-steerable states. Clearly, $\mathcal{P}_{\mathrm{ns}}^{N}$ sets up a subset of the probability majorization lattice $\mathcal{P}_n$. Moreover, by \cref{lem:prob_maj_lattice}, there exists a supremum over all non-steerable states, from which the steerability condition follows:
\begin{lemma}\label{lem:steering_ineq}
Given the orthonormal complete base sets $\mathcal{A}=(A_1,\cdots,A_N)$ and $\mathcal{B}=(B_1,\cdots,B_N)$ for Hilbert spaces $\mathcal{H}_{A}$ and $\mathcal{H}_{B}$, the violation of the following majorization inequality 
\begin{align}
\bigodot_{\mu=1}^N\vec{p}(A_\mu\otimes B_\mu)_{\rho} \prec \vec{\delta}(\mathcal{A},\mathcal{B}) \; , \; \vec{\delta}(\mathcal{A},\mathcal{B}) =\bigvee\mathcal{P}_{\mathrm{ns}}^{N}
\end{align}
signifies the steerability of the quantum state $\rho$. 
\end{lemma}
Although \cref{lem:steering_ineq} establishes a direct connection between quantum steering and probability majorization lattice, calculating the supremum $\vec{\delta}(\mathcal{A},\mathcal{B})$ presents an intractable challenge. It can be shown that in the quantum steering scenario, the supremum $\vec{\delta}(\mathcal{A},\mathcal{B})$ can be replaced by the majorization UR bound $\vec{\omega}(\mathcal{B})$, thereby formulating an operational steering detection framework. To this end, we employ the concept of \emph{aggregating a probability distribution} \cite{vidyasagar12,cicalese16,cicalese19,li20}. Given probability vectors $\vec{p}\in\mathcal{P}_n$ and $\vec{q}\in\mathcal{P}_m$, $\vec{q}$ is referred as an \emph{aggregation} of $\vec{p}$ if there is a partition $\mathcal{I}$ of $\{1,\cdots,n\}$ into disjoint sets $I_1,\cdots,I_m$ such that $q_{j}=\sum_{i\in I_j}p_i$, for $j=1,\cdots,m$. The vector $\vec{q}$ can then be denoted as $\vec{q} = \mathcal{I}(\vec{p})$ and we have the following theorem (proof shown in Supplementary Note 1):
\begin{theorem}\label{thm:steering_maj_ur}
Given orthonormal complete base sets $\mathcal{A}=(A_1,\cdots,A_N)$ and $\mathcal{B}=(B_1,\cdots,B_N)$ for Hilbert spaces $\mathcal{H}_{A}$ and $\mathcal{H}_{B}$, the violation of the following majorization inequality
\begin{align}
\bigodot_{\mu=1}^N\mathcal{I}\left(\vec{p}(\mathcal{J}(A_\mu)\otimes \mathcal{E}(B_\mu))_{\rho}\right) \prec \vec{\omega}(\mathcal{B}) 
\label{eq:steering_maj_ur}
\end{align}
signifies the steerability of quantum state $\rho$ (from Alice to Bob). Here, $\vec{\omega}(\mathcal{B})$ is the majorization UR bound of $\mathcal{B}=(B_1,\cdots,B_N)$ for binary operations $\odot\in\{\otimes,\oplus,+\}$ and $\mathcal{I}\left(\vec{p}(A_\mu\otimes B_\mu)_{\rho}\right)$ denotes all aggregations of $\vec{p}(A_\mu\otimes B_\mu)_{\rho}$ with partitions $\mathcal{I}$ satisfying $\mathcal{I}(\vec{p}\otimes\vec{q})\prec\vec{q}$; $\mathcal{E}$ denotes the local transformations preserving the majorization UR bound $\vec{\omega}(\mathcal{B})$, i.e. $\vec{\omega}(\mathcal{E}(\mathcal{B}))=\vec{\omega}(\mathcal{B})$; $\mathcal{J}$ represents any local transformation of measurement Alice performs. 
\end{theorem}

\cref{thm:steering_maj_ur} provides a criterion for quantum steering that is applicable to arbitrary finite dimension and measurement settings. It offers a novel perspective that quantum correlation information embedded in the joint probability can be extracted through appropriate aggregation operations of joint probability distributions. Next, to exhibit the capacity of the above theorem in steering detection, we first develop some steering inequalities from it and then apply the inequalities to some typical states.

\subsection*{Witnessing quantum steering via the majorization formalism}

For illustration, we herein focus on the binary operation $\oplus$ and present some practical steering inequalities. Employing the Bloch representation, a bipartite state is expressed in the form of
\begin{align}
\rho = &\frac{1}{d^2}\mathds{1}\otimes\mathds{1} + \frac{1}{2d}\vec{u}\cdot\vec{\pi}\otimes\mathds{1} + \frac{1}{2d}\mathds{1}\otimes\vec{v}\cdot\vec{\pi} \notag\\ &+ \frac{1}{4}\sum_{\mu,\nu}\mathcal{T}_{\mu\nu}\pi_{\mu}\otimes\pi_{\nu} \, .
\label{eq:bipartite_state}
\end{align}
Here, the coefficient matrix entries are
$u_{\mu}=\tr[\rho_A \pi_{\mu}]$, $v_{\mu}=\tr[\rho_B \pi_{\mu}]$ and $\mathcal{T}_{\mu\nu}=\tr[\rho \pi_{\mu}\otimes\pi_{\nu}]$ with generators $\{\pi_{\mu}\}_{\mu=1}^{d^2-1}$ of $\mathfrak{su}(d)$ Lie algebra. Let us consider the partition $\mathcal{I}=\{I_1,\cdots,I_d\}$ of $\{(1,1),\cdots,(1,n),\cdots,$ $(i,j),\cdots,(d,d)\}$ with elements
\begin{align}
I_k = \{(i,j) | j\equiv(i+k-1)\pmod d,1\leq i,j \leq d\} \; ,
\end{align}
where $k=1,\cdots,d$. It has been proved that the partition $\mathcal{I}=\{I_1,\cdots,I_d\}$ satisfies $\mathcal{I}(\vec{p}\otimes\vec{q})\prec\vec{q}$, which is equivalent to the degenerate case of Lemma 1 in Ref. \cite{guhne04}. The partition $\mathcal{I}$ results in an aggregation $\vec{\Upsilon}$ of $p(a_i,b_j|A,B;\rho)$ with components
\begin{align}
\Upsilon_k(A,B;\rho) &= \sum_{(i,j)\in I_k} p(a_i,b_j|A,B;\rho) \notag \\ 
&= \frac{1}{d} + \frac{1}{4}\sum_{(i,j)\in I_k}(\vec{a}_i,\mathcal{T}\vec{b}_j) \; ,
\end{align}
where $\vec{a}_i\left(\vec{b}_j\right)$ is Bloch vectors of projectors. Based upon \cref{thm:steering_maj_ur}, we obtain the following majorization steering inequality:
\begin{align}
\bigoplus_{\mu=1}^{N}\vec{\Upsilon}(\mathcal{J}(A_\mu),\mathcal{E}(B_{\mu});\rho) \prec\vec{\omega}(\mathcal{B}) \; ,
\end{align}
which yields a family of aggregation $\vec{\Upsilon}$ steering inequalities
\begin{align}
\mathcal{S}_L \equiv \frac{1}{L}\sum_{k=1}^{L}\left[\bigoplus_{\mu}^{N}\vec{\Upsilon}_{\mu}\right]_{k}^{\downarrow} \leq \bar{\Omega}_{L} \; , \; L=2,\cdots,Nd \; .
\label{eq:maj_steering_inequalities}
\end{align}
Here, $[\cdot]_{k}^{\downarrow}$ denotes the largest $k$-th component of a vector and $\bar{\Omega}_{L}=\frac{\Omega_{L}}{L}$ with $\Omega_{L}=\sum_{k=1}^{L}\left[\vec{\omega}(\mathcal{B})\right]_{k}^{\downarrow}$. Similarly to the qubit situation \cite{saunders10}, we define the quantity $\mathcal{S}_L$ as the steering parameter for $N$ measurement settings in arbitrary dimensional Hilbert space. The calculation of $\Omega_L$ is a typical combinatorial optimization problem (COP). We derive three computational simple upper bounds $\Lambda_L,\Theta_L$ and $\Gamma_L$ for $\Omega_L$ in the Supplementary Note 2. Especially, we have upper bounds for $\Omega_L$ in the case of MUBs:
\begin{align}
\begin{cases}
&\Theta_L = 1 + \frac{L-1}{\sqrt{d}} \; , \\
&\Gamma_L = \frac{L+\sqrt{(d-1)(dL-\Phi(L))}}{d} \; ,
\end{cases}
\label{eq:steering_approx_bound}
\end{align}
where $\Phi(L)=N\lfloor\frac{L}{N}\rfloor^2+\left(L-N\lfloor\frac{L}{N}\rfloor\right)\left(2\lfloor\frac{L}{N}\rfloor+1\right)$ with the floor function $\lfloor\cdot\rfloor$ and $1\leq L\leq N(d-1)$. Note that the upper bounds in \cref{eq:steering_approx_bound} satisfy $\Gamma_L\leq\Theta_L$ for a complete set of MUBs (see Supplementary Note 2 for proofs and details). Note that the authors of Ref. \cite{designolle19} defined the exact same quantity $\Omega_L$ and obtained the same upper bound $\Theta_L$
in terms of measurement incompatibility.

\subsection*{Applications}

Next, we formulate quantum steering inequalities in two-qubit systems, arbitrary-dimensional Werner states, isotropic states and general scenario, with detailed calculations given in Supplementary Note 3.

\emph{Two qubits system.} Given spin observables $A_{\mu}=\vec{a}_{\mu}\cdot\vec{\sigma}$ and $B_{\mu}=\vec{b}_{\mu}\cdot\vec{\sigma}$ with Pauli matrices vector $\vec{\sigma}$, we have the aggregated joint probability for a two-qubit system
\begin{align}
\Upsilon_{\pm} = \frac{1}{2}\left[1 \pm (\vec{t}\circ\vec{a}_\mu)\cdot\vec{b}_{\mu}\right] \; .
\end{align}
Here, $\vec{t}=(t_1,t_2,t_3)$ is the singular value vector of $\mathcal{T}$; $\circ$ denotes Hadamard product. Considering the two-qubit Werner state $\rho_{\mathrm{W}}=(1-w)\mathds{1}/4+w\ket{\psi_{-}}\bra{\psi_{-}}$ with $\ket{\psi_{-}} = \frac{1}{\sqrt{2}}(\ket{01} - \ket{10})$ and setting $\vec{a}_{\mu}=\vec{b}_{\mu}$, we have $\Upsilon_{\pm}=\frac{1\pm w}{2}$ and 
\begin{align}
\mathcal{S}_N = \frac{1+w}{2} \; ,
\end{align}
which yields a steering threshold for the two-qubit Werner state: $w^*=2\bar{\Omega}_{N}-1$. Throughout this paper, we define the steering threshold as the minimum value of the state parameter (e.g., the visibility $w$ for Werner states) sufficient to violate the steering inequalities under a given measurement setting. Notably, this result is equivalent to the linear steering inequality established by Saunders \emph{et al.} \cite{saunders10}. Furthermore, when measurements are conducted over the entire hemisphere of the Bloch sphere, there exists a limit of $\lim_{N\to\infty}\frac{\Omega_N}{N}=\frac{3}{4}$, which leads to a critical threshold of $1/2$. 

\begin{table}
\caption{\label{tab:steering_threshold_majo_incomp}Comparison of steering thresholds derived from the majorization lattice framework and the measurement incompatibility perspective. Note that for isotropic states, the thresholds coincide, while for Werner states, the thresholds differ.}
\renewcommand{\arraystretch}{2.0}
\setlength{\tabcolsep}{10pt} 
\begin{tabular*}{\linewidth}{@{\extracolsep{\fill}} c c c @{}}
\toprule
Quantum states & Noise robustness\cite{designolle19_robustness,designolle19,nguyen20_symmetries} & Steering threshold \\ 
\midrule
Isotropic states & $\displaystyle \frac{\lambda-N/d}{N-N/d}$ & $\displaystyle \frac{\Omega_N-N/d}{N-N/d}$ \\
Werner states  & $\displaystyle 1-\frac{d}{N}\mu$ & $\displaystyle 1-\frac{d}{N}(N-\Omega_{N(d-1)})$ \\
\bottomrule
\end{tabular*}
\end{table}

\emph{Isotropic state and Werner state in dimension $d$.} Without loss of generality, one can set $A_{\mu}=\vec{a}_{\mu}\cdot\vec{\pi}$ and $B_{\mu}=\vec{b}_{\mu}\cdot\vec{\pi}$ with $\mu=1,\cdots,N$, where $\vec{a}_\mu,\vec{b}_\mu$ are $(d^2-1)$ dimensional real vectors.
For isotropic states and Werner states, we have the aggregated joint probabilities
\begin{align}
\Upsilon_{k}^{\mathrm{ISO}}
&= \left\{
\begin{aligned}
&\frac{1+(d-1)w}{d} \; , \; k=1 \; , \\ 
&\frac{1-w}{d} \; , \; k=2,\cdots,d \; ,
\end{aligned}
\right. \\
\Upsilon_{k}^{\mathrm{W}}
&= \left\{
\begin{aligned}
&\frac{1-\eta}{d} \; , \; k=1 \; , \\ 
&\frac{d-1+\eta}{d(d-1)} \; , \; k=2,\cdots,d \; .
\end{aligned}
\right.
\end{align}
and steering parameters
\begin{align}
\mathcal{S}_N^{\mathrm{ISO}} = \frac{1+(d-1)w}{d} \; , \; \mathcal{S}_{N(d-1)}^{\mathrm{W}} = \frac{d-1+\eta}{d(d-1)} \; .
\end{align}
Here, $w$ and $\eta$ $\in [0,1]$ are noise parameters for the isotropic and Werner state, respectively. By \cref{eq:maj_steering_inequalities}, we obtain the steering thresholds for isotropic and Werner states, respectively
\begin{align}
w^* &=\left(d\bar{\Omega}_{N}-1\right)/(d-1) \; , \label{eq:isotropic_threshold} \\ 
\eta^* &=(d-1)\left(d\bar{\Omega}_{N(d-1)}-1\right) \; . \label{eq:werner_threshold} 
\end{align}
For the case of two measurement settings, we obtain $\bar{\Omega}_2=(1+c)/2$ with $c=\max_{i,j}|\braket{a_i|b_j}|$ representing the maximal overlap of the two bases. The steering threshold is then $w^*=\frac{d(1+c)-2}{2(d-1)}$,
which implies that the results in Refs. \cite{li15,zeng18,guo19} are the special cases corresponding to pairs of MUBs. A notable correspondence exists between the steering thresholds derived from the majorization lattice framework and those obtained from the measurement incompatibility perspective \cite{designolle19_robustness,designolle19,nguyen20_symmetries}. This relationship becomes apparent when reformulating \cref{eq:isotropic_threshold,eq:werner_threshold} in terms of $\Omega_N$ and $\Omega_{N(d-1)}$, as shown in \cref{tab:steering_threshold_majo_incomp}. Remarkably, for the isotropic states, our result coincides exactly with the noise robustness derived from the perspective of measurement incompatibility \cite{designolle19_robustness,nguyen20_symmetries}.

\begin{figure}
\centering
\includegraphics[width=0.5\textwidth]{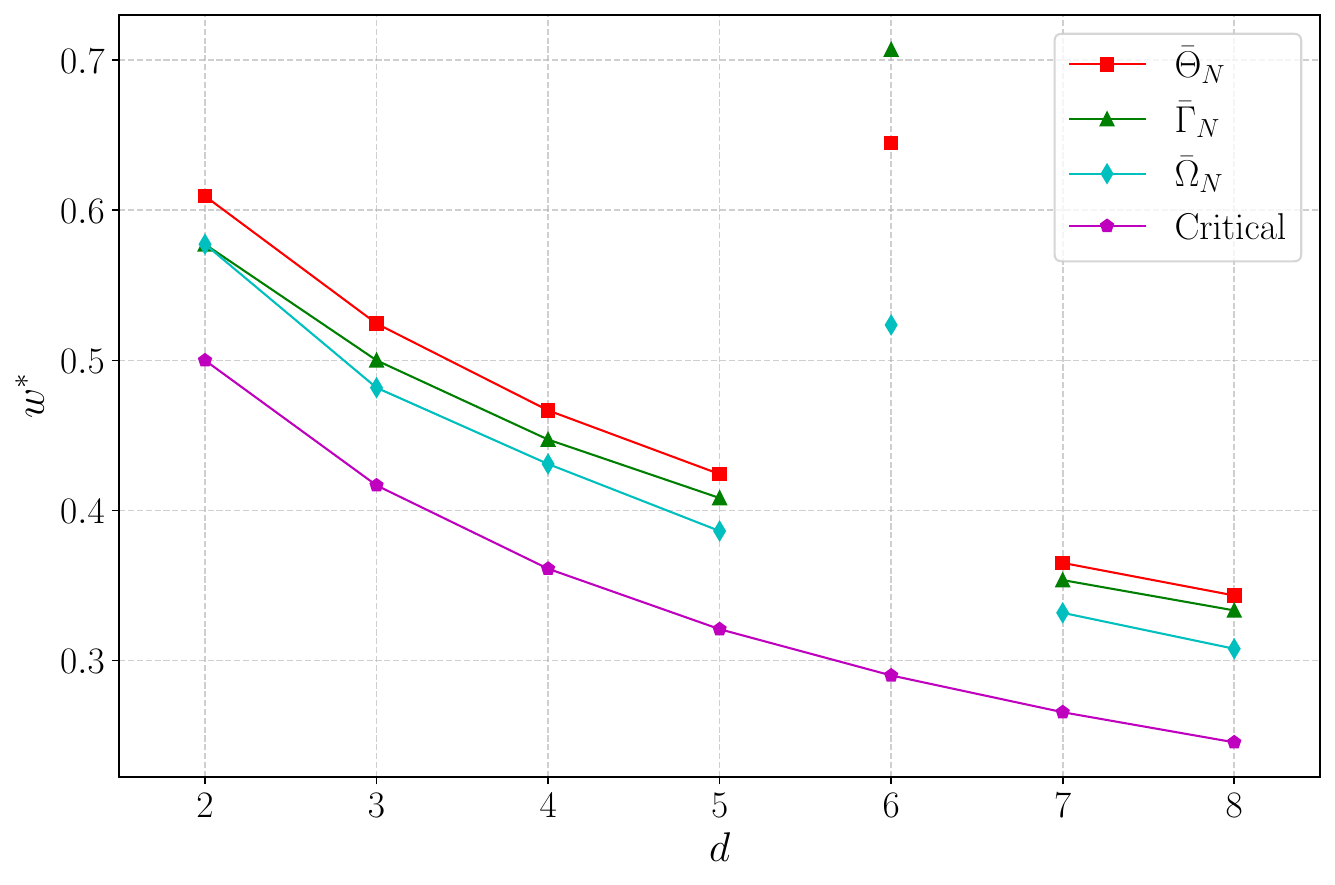}
\caption{The steering thresholds of the isotropic states for different dimensions $d$, in case of a complete set of MUBs. The red and green lines represent the steering threshold derived from the approximate upper bounds $\bar{\Theta}_N$ and $\bar{\Gamma}_N$, respectively. The cyan line indicates the optimal steering threshold from $\bar{\Omega}_N$, while the magenta line represents the critical value of isotropic states derived from the LHS model \cite{werner89,wiseman07,almeida07}. Note that, due to the existence problem of $6$-dimensional MUBs, the threshold for $d=6$ is calculated using only three MUBs.}
\label{fig:isotropic_mubs_steering}
\end{figure}

\begin{table}
\caption{\label{tab:steering_threshold_approx_bound}Steering thresholds of isotropic ($w^*$) and Werner ($\eta^*$) states derived from upper bounds $\bar{\Gamma}_N$ and $\bar{\Theta}_N$.}
\renewcommand{\arraystretch}{2.2}
\begin{tabular*}{\hsize}{@{}@{\extracolsep{\fill}} l c c @{}}
\toprule
Bounds & $w^*$ & $\eta^*$ \\ 
\midrule
$\bar{\Gamma}_N$ & $\displaystyle \frac{1}{\sqrt{N}}$ & $\displaystyle \frac{d-1}{\sqrt{N}}$ \\
$\bar{\Theta}_N$ & $\displaystyle \frac{(N+\sqrt{d})(\sqrt{d}-1)}{N(d-1)}$ & $\displaystyle \frac{(\sqrt{d}-1) ((d-1) N+\sqrt{d})}{N}$ \\
\bottomrule
\end{tabular*}
\end{table}
Utilizing upper bounds derived in \cref{eq:steering_approx_bound}, we obtain approximate steering thresholds for isotropic states and Werner states, as summarized in \cref{tab:steering_threshold_approx_bound}. These bounds provide an excellent approximation for isotropic state steering scenario involving MUBs, which significantly simplifies the analysis of steering in high-dimensional system. Note, the thresholds in Refs. \cite{skrzypczyk15,qu22,costa18} are compatible with the upper bounds of $\bar{\Theta}_N$ and $\bar{\Gamma}_N$, respectively, which indicates that those previous results are not optimal ones for isotropic state. As illustrated in \cref{fig:isotropic_mubs_steering}, we compute the optimal steering thresholds for various isotropic states with MUBs dimensions, from $d=2$ to $8$ for illustration, and compare these thresholds with those from $\bar{\Gamma}_N,\bar{\Theta}_N$, as well as the critical values from LHS model \cite{werner89,wiseman07,almeida07}. 

It is noteworthy that for $d>2$, the steering threshold of Werner states in \cref{tab:steering_threshold_approx_bound} exceeds $1$, i.e. $\eta^*\geq 1$, which indicates that MUBs fail to witness the steerability of Werner states in high dimensions. Intuitively, this follows from $\eta^*=1-\frac{d}{N}(N-\Omega_{N(d-1)})$, where $\Omega_{N(d-1)}$ is close to $N$ for MUBs. Conversely, non-MUB measurements with $\Omega_{N(d-1)}<N$ are expected to perform better. It is worth noting that Nguyen \textit{et al.} showed that MUBs can never detect steering for Werner states in odd prime-power dimensions, a limitation that appears to hold asymptotically for even prime-power dimensions (e.g. $\eta^*=0.9174$ for $d=4$) \cite{nguyen20_symmetries}. These findings highlight an intriguing fact that while MUBs effectively capture the steerability of isotropic states, they perform poorly for witnessing the steerability of Werner states in high dimensions.
\begin{figure*}[ht]
\subfloat[Qutrit isotropic states]{\label{fig:qutrit_isotropic_steering}\includegraphics[width=0.5\textwidth]{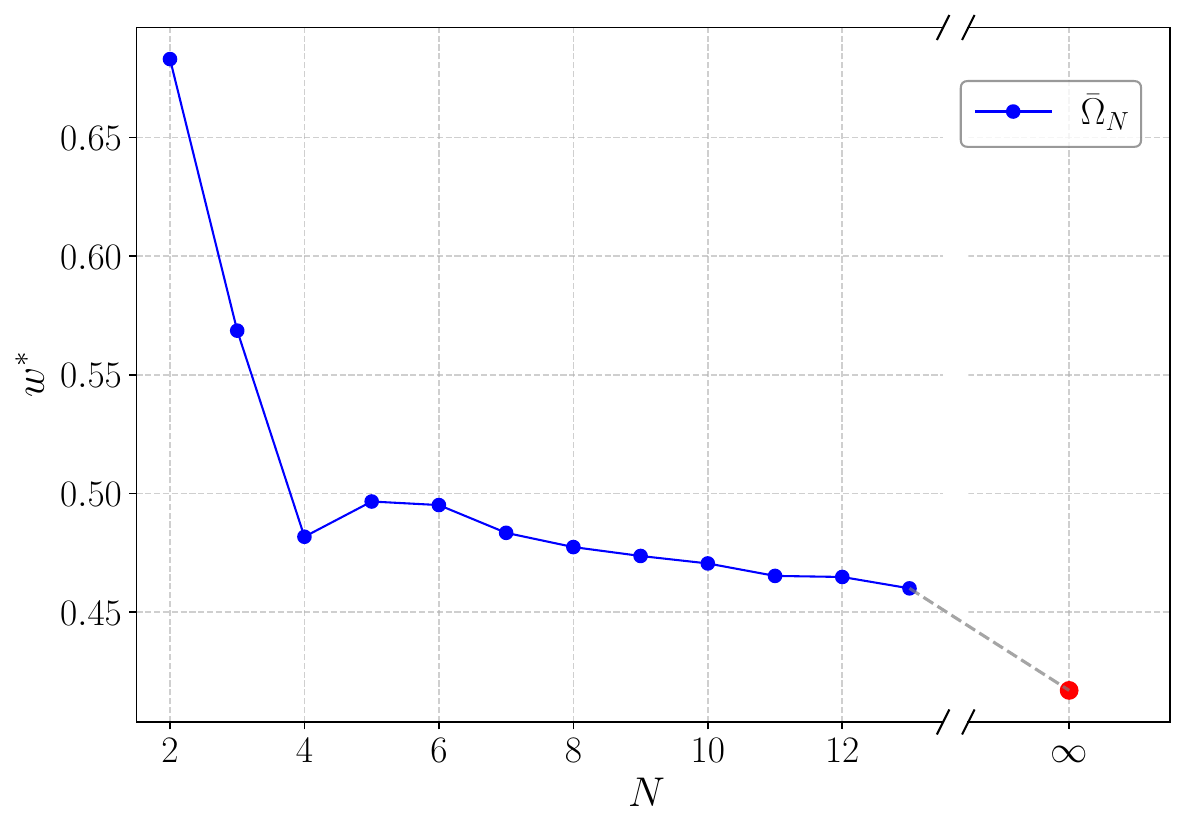}}
\subfloat[Qutrit Werner states]{\label{fig:qutrit_werner_steering}\includegraphics[width=0.5\textwidth]{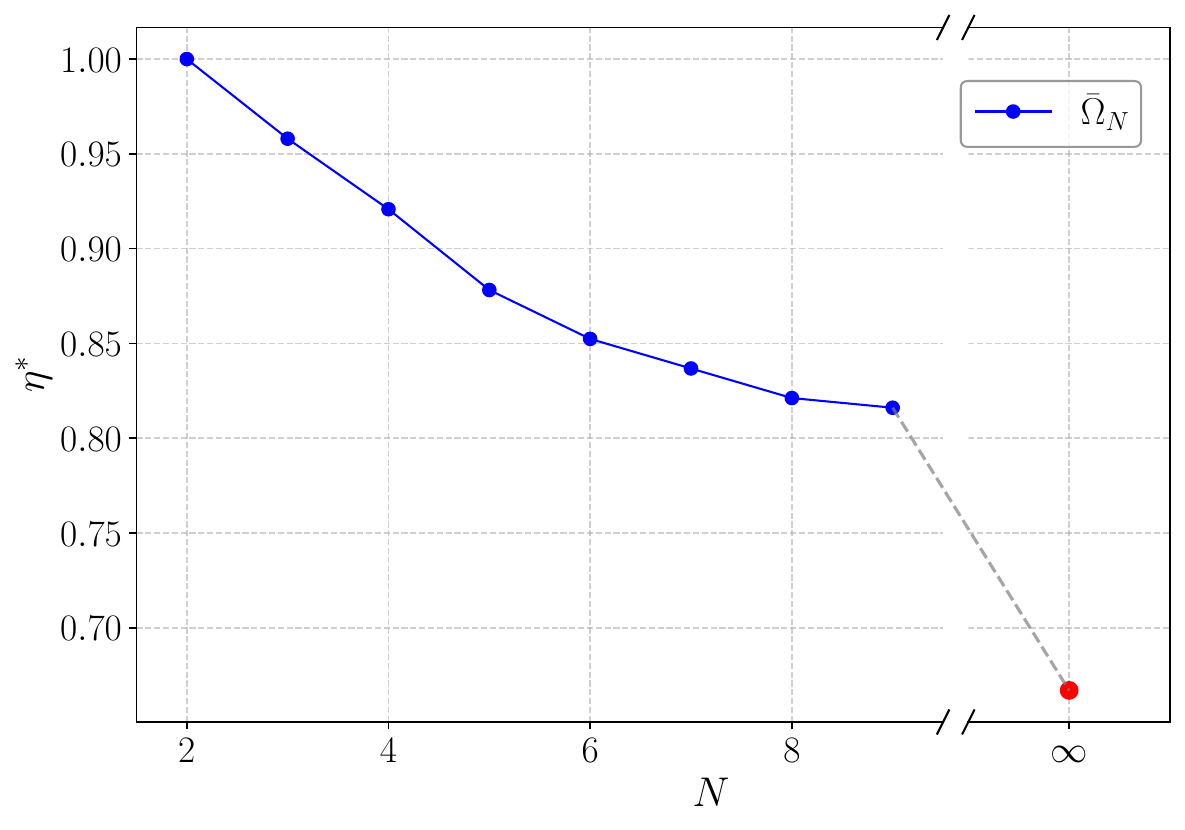}}
\caption{
The steering threshold value of the qutrit isotropic and Werner states for $N$ measurement settings. Here, we plot the steering threshold of qutrit isotropic states (a) and Werner states (b) obtained from the Cross-Entropy Method (CEM) for $N=2$ to $13$ and $N=2$ to $8$, respectively. The red points represent the critical values $5/12$ and $2/3$ of qutrit isotropic and Werner states from LHS model \cite{werner89,wiseman07,almeida07}. 
}
\label{fig:qutrit_steering}
\end{figure*}

Notice that the new approach enables the investigation of optimal measurement settings for quantum steering. Consequently, we are tasked with addressing the optimization problem
\begin{align}\label{eq:steering_opt_problem}
\textrm{min}\quad &\bar{\Omega}_N \\
\textrm{s.t.} \quad &\mathcal{B}=(B_1,\cdots,B_N)\; , \forall B_\mu=B_\mu^\dagger \; . \notag 
\end{align}
In \cref{fig:qutrit_steering}, we examine the optimal settings for qutrit isotropic and Werner states using the Cross-Entropy Method (CEM) \cite{deboer05,botev13}, a powerful technique for solving complex continuous optimization problems and machine learning tasks. As shown in \cref{fig:qutrit_isotropic_steering,fig:qutrit_werner_steering}, we compute the steering thresholds for qutrit isotropic states in $N=2$ to $13$ measurement settings and for qutrit Werner states in $N=2$ to $8$ measurement settings. Our results indicate that MUBs remain optimal for $N=2,3,4$ in isotropic scenario. Notably, Bavaresco \textit{et al.} obtained smaller upper bounds for projective measurements using semidefinite programming (SDP) and search algorithms \cite{bavaresco17}, resulting in a small gap between their results and our lower bounds. A plausible explanation is that both the CEM and the search algorithms are heuristic optimization methods, which do not guarantee finding the global optimum. To certify the steerability of qutrit Werner states, the optimal measurement settings always be non-MUBs. Lower thresholds are achieved as the number of measurement settings $N$ increases. We conjecture that the critical values $5/12$ and $2/3$ can be attained in the limit as $N\to\infty$ for both qutrit isotropic and Werner states, respectively, similarly to the scenario in two-qubit Werner states.

\emph{General scenario.} Here, we formulate the steering detection in a general scenario where Alice's measurements are related to Bob's measurements via unitary or anti-unitary transformations, i.e., $A_{\mu}=\mathcal{J}(B_{\mu})$. This leads to the following steering inequality:
\begin{align}
\bigoplus_{\mu=1}^{N}\vec{\Upsilon}(\mathcal{J}(B_\mu),B_{\mu};\rho) \prec\vec{\omega}(\mathcal{B}) \; .
\label{eq:majo_steering_ineq_gene}
\end{align}
Since the joint probability distributions $\vec{\Upsilon}$ depend on the choice of $\mathcal{J}$, there exists an optimal alignment between Alice's and Bob's measurements for steering detection. The optimal alignment mechanism necessarily minimizes the entropy of the aggregated joint probability distributions. Therefore, the optimal $\mathcal{J}$ can be obtained by solving the optimization problem:
\begin{align}
\textrm{min}\quad &H\left(\bigoplus_{\mu=1}^{N}\vec{\Upsilon}(\mathcal{J}(B_\mu),B_{\mu};\rho)\right) \\
\textrm{s.t.} \quad & \mathcal{J} \text{ is unitary or anti-unitary} \; , \notag 
\end{align}
where $H(\cdot)$ denotes the Shannon entropy.

\begin{figure*}[ht]
\subfloat[Steering of state $\rho(\lambda,\theta,\phi)$ with MUBs]{\label{fig:qutrit_steering_mubs}\includegraphics[width=0.5\textwidth]{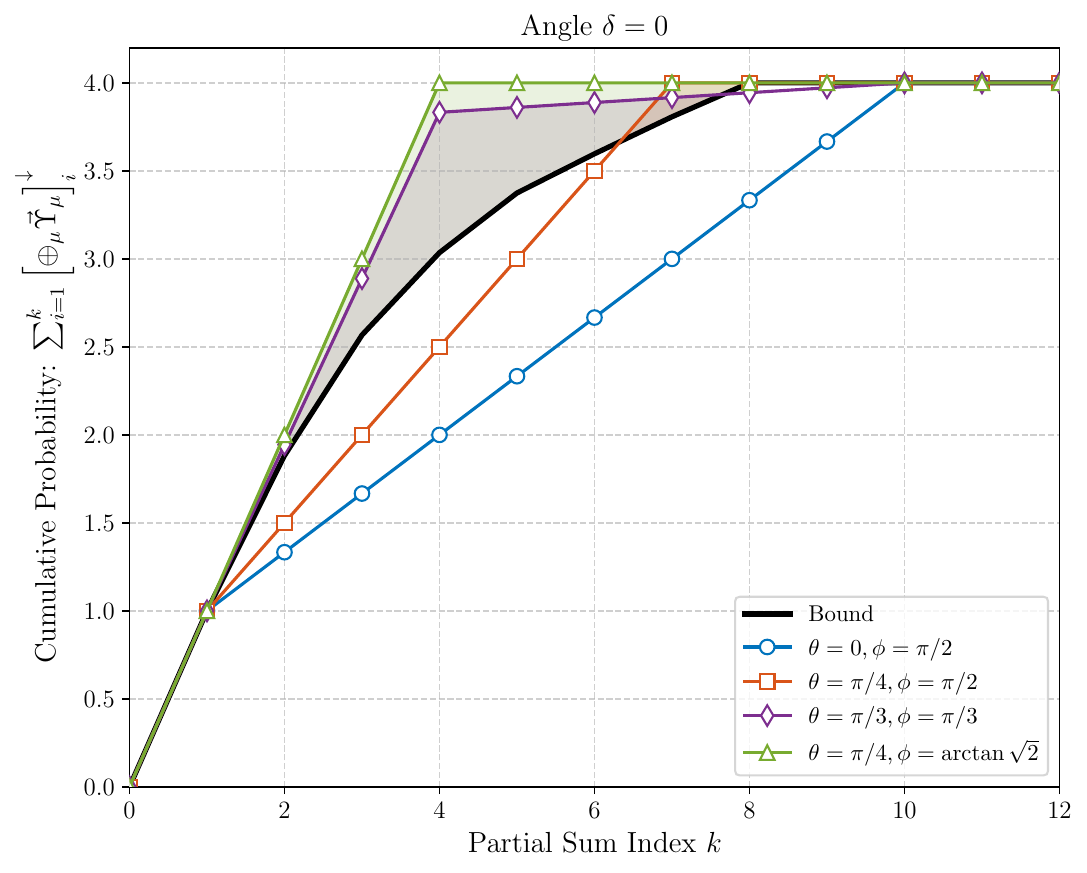}}
\subfloat[Steering of state $\rho(\lambda,\theta,\phi)$ with non-MUBs]{\label{fig:qutrit_steering_non-mubs}\includegraphics[width=0.5\textwidth]{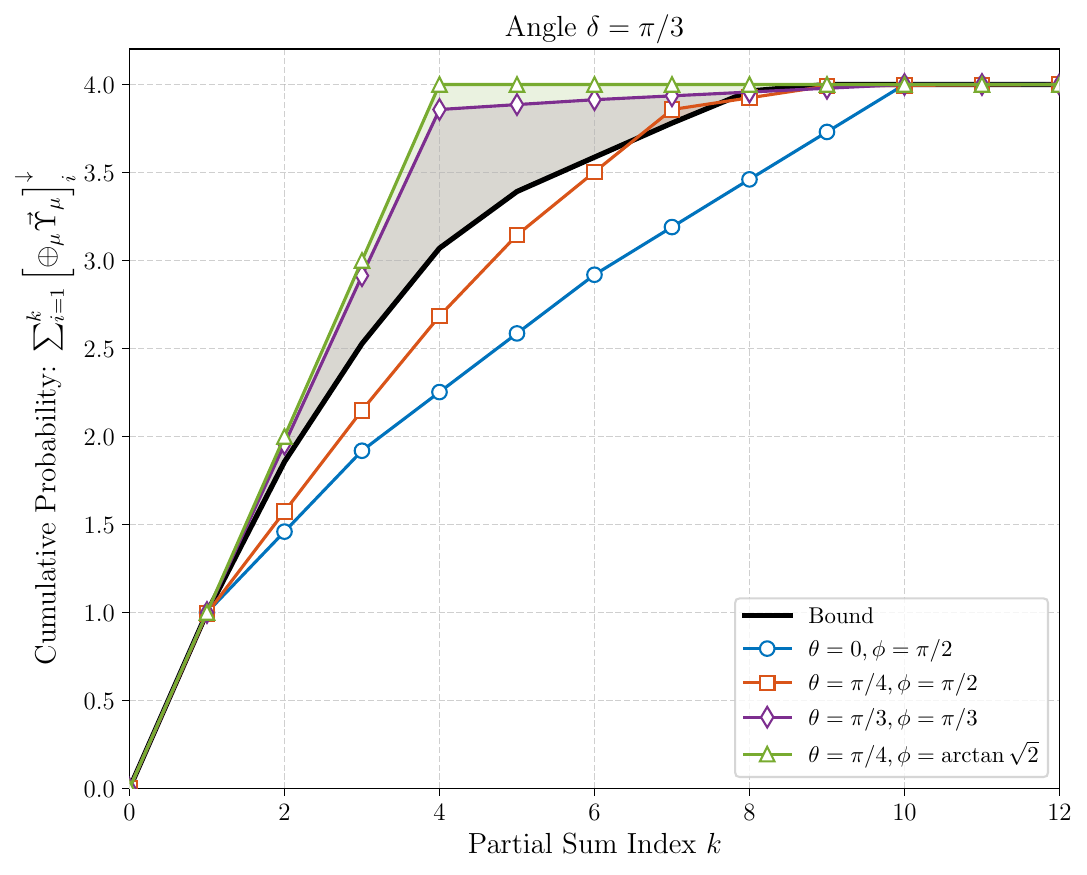}}
\caption{Lorenz curves for $\rho(\lambda,\theta,\phi)$ with Schmidt rank $1(\theta=0,\phi=\pi/2)$, $2(\theta=\pi/4,\phi=\pi/2)$ and $3(\theta=\phi=\pi/3,\theta=\pi/4,\phi=\arctan\sqrt{2})$ for MUBs and non-MUBs cases. The black solid line represents the majorization bound $\vec{\omega}(\mathcal{B})$. The shadow region above the majorization bound curve indicates the steerable region detected by \cref{eq:majo_steering_ineq_gene}.
}
\label{fig:qutrit_steering_general}
\end{figure*}

We now apply the above procedure to detect steerability in a general scenario. Consider the following family of qutrit states:
\begin{align}
\rho(\lambda,\theta,\phi) = \lambda\ket{\psi(\theta,\phi)}\bra{\psi(\theta,\phi)} + (1-\lambda)\frac{\mathds{1}}{9} \; ,
\end{align}
where $\ket{\psi(\theta,\phi)}=\cos\theta\sin\phi\ket{00} + \sin\theta\sin\phi\ket{11} + \cos\phi\ket{22}$, with $\lambda\in[0,1]$, $\theta\in[0,\pi/4]$ and $\phi\in[0,\pi/2]$. This state reduces to the isotropic state when $\theta=\pi/4$ and $\phi=\arctan\sqrt{2}$. We consider the measurement setting $\mathcal{B}=(B_1,B_2,UB_3U^\dagger,UB_4U^\dagger)$ with $U=e^{i\delta J_z}$, where $B_1,B_2,B_3,B_4$ are MUBs and $J_z$ is the $z$ component of the angular momentum operator. 
In \cref{fig:qutrit_steering_general}, we plot the Lorenz curves of \cref{eq:majo_steering_ineq_gene} for $\rho(\lambda,\theta,\phi)$ with Schmidt rank 1 ($\theta=0,\phi=\pi/2$), rank 2 ($\theta=\pi/4,\phi=\pi/2$), and rank 3 ($\theta=\phi=\pi/3$, $\theta=\pi/4,\phi=\arctan\sqrt{2}$) for both MUBs (\cref{fig:qutrit_steering_mubs}) and non-MUBs (\cref{fig:qutrit_steering_non-mubs}) cases. The Lorenz curve of a probability distribution vector $\vec{p}$ is defined as $f_{\vec{p}}(k)\equiv\sum_{i=1}^{k}p_i^{\downarrow}$ with $f_{\vec{p}}(0)=0$. Note that Lorenz curves have been previously applied to investigate optimal uncertainty relations \cite{li19}. Genuine high-dimensional steering has attracted extensive attention recently \cite{designolle21,de23_steering,dalessandro25}. These studies show that high-dimensional entanglement, as quantified by the Schmidt number, can lead to a stronger form of steering that is provably impossible to obtain via entanglement in lower dimensions. The results in \cref{fig:qutrit_steering_general} and \cref{tab:steering_noise_threshold} indicate that states with higher Schmidt rank reveal stronger robustness to noise (corresponding to a larger shaded area above the majorization bound curve). Conversely, lower Schmidt rank states (e.g., $\theta=\pi/4,\phi=\pi/2$) exhibit more pronounced sensitivity to noise and the choice of measurement settings.
\begin{table}
\caption{\label{tab:steering_noise_threshold}Noise threshold $\lambda^*$ of $\ket{\psi(\theta,\phi)}$ for MUBs ($\delta=0$) and non-MUBs ($\delta=\pi/3$) cases. }
\centering
\renewcommand{\arraystretch}{1.5} 
\setlength{\tabcolsep}{4pt}      

\begin{tabular*}{\linewidth}{@{\extracolsep{\fill}} c c c c c @{}}
\toprule
$\delta$ & \makecell{$\theta=0$\\$\phi=\pi/2$} 
         & \makecell{$\theta=\pi/4$\\$\phi=\pi/2$} 
         & \makecell{$\theta=\pi/3$\\$\phi=\pi/3$} 
         & \makecell{$\theta=\pi/4$\\$\phi=\arctan\sqrt{2}$} \\ 
\midrule
$0$      & 1 & $0.8417$ & $0.5140$ & $0.4818$ \\
$\pi/3$  & 1 & $0.9494$ & $0.6879$ & $0.6512$ \\
\bottomrule
\end{tabular*}
\end{table}

\section*{Discussion}

In this paper, we have established a connection between quantum steering and the probability majorization lattice, leading to a novel framework for detecting quantum steering. This framework is applicable to arbitrary finite dimensions and measurement settings. By introducing the concept of aggregating probability distributions, we have formulated a family of aggregation-based steering inequalities and applied them to the two-qubit systems, arbitrary-dimensional Werner and isotropic states and general scenarios.

From the perspective of information theory, our approach facilitates the extraction of quantum correlation information embedded in the joint probability without loss, utilizing appropriate aggregation operations that differ from those based on entropies and variances. In the context of mutually unbiased bases (MUBs), our approach not only reproduces known steering inequalities as approximate cases but also provides improved thresholds and new insights into the optimal measurement settings for detecting quantum steerability.

For high-dimensional Werner states, non-MUB measurements outperform MUBs in witnessing steerability. We investigate the optimal $N$-measurement settings for qutrit isotropic and Werner states using the Cross-Entropy Method (CEM), revealing that MUBs remain optimal for $N=2,3,4$ for isotropic states, while non-MUBs are required for qutrit Werner states. Our framework can also be employed to explore quantum steering in more general scenarios by optimizing the alignment between Alice's and Bob's measurements. We illustrate this by examining a family of non-symmetric qutrit states.

High-dimensional quantum systems have garnered significant attentions in quantum information processing due to their enhanced resilience to noise, reduced susceptibility to loss, and greater information capacity. These advantages render them particularly valuable for applications in quantum communication \cite{branciard12,islam17,cozzolino19}, quantum cryptography \cite{groblacher06}, and superdense coding \cite{hu18_ququarts}. Hopefully, the majorization-based framework for quantum steering may enlighten our understanding of quantum correlation and facilitate practical applications across these areas.

\section*{Methods}

\subsection*{$\Omega_L$ of $N$ measurement bases}\label{appen:Omega_N_definition}

Given $N$ orthonormal and complete bases $\{\ket{b}_i^{(\mu)}\}_{i=1}^d$ for $\mu=1,2,\cdots,N$, we define the index set $I=\{1,2,\cdots,d\}$ and the subsets $I_1,I_2,\cdots,I_N\subset I$, where $I_\mu$ denotes the index set of the $\mu$-th basis $A_\mu$. 
Assuming we choose $L$ base vectors from the $N$ bases, with each basis selecting $|I_\mu|$ vectors, we can define $L$ projectors $\{\ket{x_j}\bra{x_j}\}$, $j=1,2,\cdots,L=\sum_{\mu=1}^{N}|I_\mu|$ as follows:
\begin{align}
&\ket{x_{i_1}} = \ket{a_i^{(1)}} \; , \; \exists i\in I_1 \; , \; i_1=1,\cdots,|I_1| \; , \\ 
&\ket{x_{i_2}} = \ket{a_{i}^{(2)}} \; , \; \exists i\in I_2 \; , \; i_2=|I_1|+1,\cdots,|I_1|+|I_2| \; , \\ 
&\hspace{6ex}\vdots \notag \\
&\ket{x_{i_N}} = \ket{a_{i}^{(N)}} \; , \; \exists i\in I_N \; , \; i_N=\sum_{\mu=1}^{N-1}|I_\mu|+1,\cdots,L=\sum_{\mu=1}^{N}|I_\mu| \; .
\end{align}
Then, we can reformulate $\Omega_L$ as
\begin{align}
\Omega_{L}
&\equiv\max_{\substack{I_1,\cdots,I_N \\ \sum_{\mu=1}^{N}|I_\mu|=L}}\max_{\rho}\tr\left[\rho\left(\sum_{j=1}^L\ket{x_j}\bra{x_j}\right)\right] \\ 
&= \max_{\substack{I_1,\cdots,I_N \\ \sum_{\mu=1}^{N}|I_\mu|=L}}\left\|\sum_{j=1}^L\ket{x_j}\bra{x_j}\right\| \; .
\end{align}
Here, $\|\cdot\|$ denotes the operator norm, i.e. the largest singular value of the operator. Especially, if $L>d(N-1)$ we have $\Omega_L=N$ (the maximal value possible), which occurs if $I_1=\cdots=I_{N-1}=I$ and $I_N=\{l\}$, $\rho=\ket{a_l^{(N)}}\bra{a_l^{(N)}}$. The solution of $\Omega_L$ is a typical combinatorial optimization problem (COP), which has a discrete set of feasible solutions with a size of $\binom{Nd}{L}$. Next, we provide a few equivalent expressions of $\Omega_L$ for $N$ measurements.

\subsection*{Bloch representation of $\Omega_L$}
Let $\{\vec{x}_j\}$ be Bloch vectors of $\{\ket{x_j}\}$. Under Bloch representation, the projectors are in form of $\ket{x_j}\bra{x_j}=\frac{1}{d}\mathds{1}+\frac{1}{2}\vec{x}_j\cdot\vec{\pi}$. Thus, we have
\begin{align}
\Omega_{L} &= \max_{\substack{I_1,\cdots,I_N \\ \sum_{\mu=1}^{N}|I_\mu|=L}}\max_{\rho}\tr\left[\rho\left(\sum_{j=1}^L\ket{x_j}\bra{x_j}\right)\right] \\
&= \max_{\substack{I_1,\cdots,I_N \\ \sum_{\mu=1}^{N}|I_\mu|=L}}\max_{\rho}\tr\left[\rho\left(\frac{L}{d}\mathds{1}+\vec{o}_L\cdot\vec{\pi}\right)\right] \\
&=\frac{L}{d} + \max_{\substack{I_1,\cdots,I_N \\ \sum_{\mu=1}^{N}|I_\mu|=L}}\left\|\vec{o}_L\cdot\vec{\pi}\right\|  \; .
\end{align}
Here, $\vec{o}_L=\frac{1}{2}\sum_{j=1}^L\vec{x}_j$. Clearly, a concise expression can be derived for the qubit system
\begin{align}
\Omega_{L} = \frac{L}{2} + \max_{\substack{I_1,\cdots,I_N \\ \sum_{\mu=1}^{N}|I_\mu|=L}}\left|\vec{o}_L\right| \; .
\label{eq:steering_omega_bloch_qubit}
\end{align}

\subsection*{Transition matrix representation} Here, we formulate $\Omega_L$ via the transition matrices between measurement bases.
\begin{align}
\Omega_{L}
&= \max_{\substack{I_1,\cdots,I_N \\ \sum_{\mu=1}^{N}|I_\mu|=L}}\max_{\rho}\tr\left[\rho\left(\sum_{j=1}^L\ket{x_j}\bra{x_j}\right)\right] \; , \\ 
&= \max_{\substack{I_1,\cdots,I_N \\ \sum_{\mu=1}^{N}|I_\mu|=L}}\max_{\ket{\psi}}\sum_{j=1}^L|\braket{x_j|\psi}|^2 \; , \\ 
&= \max_{\substack{I_1,\cdots,I_N \\ \sum_{\mu=1}^{N}|I_\mu|=L}}\max_{\ket{\psi}}\braket{\psi|X^\dagger X|\psi} \; , \\
&= \max_{\substack{I_1,\cdots,I_N \\ \sum_{\mu=1}^{N}|I_\mu|=L}}\lambda_{\max}(\mathcal{X}) \; .
\end{align}
Here, we have employed the property of operator norm in the last line and $\lambda_{\max}(\mathcal{X})$ denotes the largest eigenvalue of the operator $\mathcal{X}=XX^\dagger$ and $X=\left(\ket{x_1},\ket{x_2},\cdots,\ket{x_k}\right)$. The operator $\mathcal{X}$ is defined as 
\begin{align}
\mathcal{X}=XX^\dagger 
&= \left[
\begin{matrix}
\mathds{1}_{|I_1|} & U_{(1,2)}[I_1,I_2] & \cdots & U_{(1,N)}[I_1,I_N] \\ 
U_{(2,1)}^{\dagger}[I_2,I_1] & \mathds{1}_{|I_2|} & \cdots & U_{(2,N)}[I_2,I_N] \\ 
\vdots & \vdots & \ddots & \vdots \\
U_{(N,1)}^{\dagger}[I_N,I_1] & U_{(N,2)}^{\dagger}[I_N,I_2] & \cdots & \mathds{1}_{|I_N|}
\end{matrix}
\right] \; ,
\end{align}
where $\mathds{1}_{|I_\mu|}$ is the identity matrix of size $|I_\mu|\times |I_\mu|$ and $U_{(i,j)}[I_\mu,I_\nu]$ is the submatrix of entries that lie in the rows of the transition matrix $U_{(i,j)}=\{\braket{a_i^{(\mu)}|a_j^{(\nu)}}\}_{i,j=1}^{d}$ indexed by $I_\mu$ and the columns indexed by $I_\nu$, i.e. $U_{(i,j)}[I_\mu,I_\nu]=\{\braket{a_i^{(\mu)}|a_j^{(\nu)}}\}_{i\in I_\mu,j\in I_\nu}$.

\emph{Two measurement bases scenario.} When the involved measurement bases are two orthonormal complete bases, i.e. $N=2$, we have 
\begin{align}
\mathcal{X}&= \left[
\begin{matrix}
\mathds{1}_{|I_1|} & U[I_1,I_2] \\ 
U^{\dagger}[I_2,I_1] & \mathds{1}_{|I_2|} 
\end{matrix}
\right] = \mathds{1}_{|I_1|+|I_2|} + \left[
\begin{matrix}
0 & U[I_1,I_2] \\ 
U^{\dagger}[I_2,I_1] & 0 
\end{matrix}
\right]
\end{align}
In light of Theorem 7.3.3 in Ref. \cite{horn13-svd}, we have $\lambda_{\max}(\mathcal{X})=1+\sigma_{\max}(U[I_1,I_2])$, and thus
\begin{align}
\Omega_{L} = 1+\max_{\substack{I_1,I_2 \\ |I_1|+|I_2|=L}}\sigma_{\max}(U[I_1,I_2]) \; , \; L\leq d \; , \\ 
\Omega_{L} = 2 \; , \; L> d \; .
\end{align}
Here, $\sigma_{\max}(\cdot)$ denotes the maximal singular value and $U[I_1,I_2]=\{u_{ij}\}_{i\in I_1,j\in I_2 }$ for $u_{ij}=\braket{a_i^{(1)}|a_j^{(2)}}$. If $|I_1|=1$ or $|I_2|=1$, we have $\sigma_{\max}(U[I_1,I_2])=\sqrt{\sum_{j\in I_n}|u_{ij}|^2}$ or $\sigma_{\max}(U[I_m,I_1])=\sqrt{\sum_{i\in I_m}|u_{ij}|^2}$. So, the first three terms have the analytical expressions
\begin{align}
&\Omega_1 = 1 \; , \; \Omega_2 = 1 + \max_{i,j}|\braket{a_i^{(1)}|a_j^{(2)}}| \; , \; \\
&\Omega_3 = 1 + \max_{\substack{i= i',j\neq j' \\ j=j',i\neq i'}}\sqrt{|\braket{a_i^{(1)}|a_j^{(2)}}|^2+|\braket{a_{i'}^{(1)}|a_{j'}^{(2)}}|^2} \; .
\end{align}

\emph{Mutually unbiased bases.} For a pair of mutually unbiased bases (MUBs), the transition matrix $U$ between them is a discrete Fourier transformation (DFT) matrix, i.e.
\begin{align}
U = 
\frac{1}{\sqrt{d}}\left[\begin{array}{ccccc}
1 & 1 & 1 & \cdots & 1 \\
1 & \omega & \omega^2 & \cdots & \omega^{d-1} \\
1 & \omega^2 & \omega^4 & \cdots & \omega^{2(d-1)} \\
1 & \omega^3 & \omega^6 & \cdots & \omega^{3(d-1)} \\
\vdots & \vdots & \vdots & \ddots & \vdots \\
1 & \omega^{d-1} & \omega^{2(d-1)} & \cdots & \omega^{(d-1)(d-1)}
\end{array}\right] \; ,
\end{align}
where $\omega=e^{2\pi i/d}$ is a $d$-th root of unity. For a pair of $d$-dimensional MUBs, we conjecture the following expression. It is noted that the matrix $U$ is only one possible example of pairs of MUBs. There exist inequivalent pairs of MUBs \cite{brierley10,designolle19_robustness}.

\emph{Conjecture.}
\begin{align}
\Omega_{L} = 1+\max_{\substack{I_m,I_n \\ |I_m|=\lfloor\frac{L}{2}\rfloor,|I_n|=\lceil\frac{L}{2}\rceil}}\sigma_{\max}(U[I_m,I_n]) \; , \; L\leq d \; .
\end{align}
Here, $\lfloor\cdot\rfloor$ and $\lceil\cdot\rceil$ denote the floor and the ceiling functions, respectively.

\section*{Data Availability}

All relevant data used for Examples and Figs. are available from the authors.

\section*{Code Availability}

The codes used to generate the results presented in this paper are available at the repository  \url{https://github.com/yangmacheng/High-Dimensional-Quantum-Steering}.

\section*{Acknowledgements}

This work was supported in part by the National Natural Science Foundation of China (NSFC) under the Grants 12475087, 12235008, the Fundamental Research Funds for Central Universities, and China Postdoctoral Science Foundation funded project No. 2024M753174.

\section*{Author Contributions}

All authors have equally contributed to the main result, the examples and the writing. All authors have given approval for the final version of the manuscript.

\section*{Competing Interests}

The authors declare no competing interests.

\section*{Additional Information}

Correspondence and requests for materials should be addressed to Ma-Cheng Yang or Cong-Feng Qiao.


\newpage
\appendix
\setcounter{equation}{0}
\setcounter{figure}{0}
\setcounter{table}{0}
\setcounter{section}{0}
\renewcommand{\theequation}{S\arabic{equation}}
\renewcommand{\bibnumfmt}[1]{[S#1]}
\renewcommand{\citenumfont}[1]{S#1}
\renewcommand{\thesection}{Supplementary Note \arabic{section}:}
\renewcommand{\refname}{Supplementary References}
\renewcommand{\figurename}{Supplementary Figure}
\renewcommand{\tablename}{Supplementary Table}

\newcommand{\suppfigref}[1]{Supplementary Figure~\ref{#1}}
\newcommand{\supptabref}[1]{Supplementary Table~\ref{#1}}

\renewcommand{\thefootnote}{\fnsymbol{footnote}}

\begin{center}
\textbf{\large Witness High-Dimensional Quantum Steering via Majorization Lattice \\ \vspace{1ex} Supplemental Material} \\
\author
{Ma-Cheng Yang$^{1}$ and Cong-Feng Qiao$^{1,2}$\footnote[2]{correspondence: \email{qiaocf@ucas.ac.cn}}\\
\normalsize{$^{1}$School of Physical Sciences, University of Chinese Academy of Sciences}\\
\normalsize{No.1 Yanqihu East Rd, Beijing 101408, China}\\
\normalsize{$^{2}$Key Laboratory of Vacuum Physics, University of Chinese Academy of Sciences}\\
\vspace{-1.5ex}
\normalsize{No.1 Yanqihu East Rd, Beijing 101408, China}
}
\end{center}

\section{The proof of Theorem 1}\label{appen:theorem1_steering_majorization}

Given $\vec{p}=(p_1,\cdots,p_n)\in\mathcal{P}_n$, we say that $\vec{q}=(q_1,\cdots,q_m)\in\mathcal{P}_m$ is an \emph{aggregation} of $\vec{p}$ if there is a partition $\mathcal{I}$ of $\{1,\cdots,n\}$ into disjoint sets $I_1,\cdots,I_m$ such that $q_{j}=\sum_{i\in I_j}p_i$, for $j=1,\cdots,m$, simply denoted as $\vec{q}=\mathcal{I}(\vec{p})$ with $\mathcal{I}=\{I_1,\cdots,I_m\}$. The aggregations of a probability distribution contain the less uncertainty than the original one, and there is the following result \cite{cicalese16}:
\begin{lemma}
Given $\vec{p}\in\mathcal{P}_n$ and any aggregation $\vec{q}\in\mathcal{P}_m$ of $\vec{p}$, we have $\vec{p}\prec\vec{q}$.
\end{lemma}
Let $\vec{p}\in\mathcal{P}_n$ and $\vec{q}\in\mathcal{P}_m$ be any two probability distributions. We construct a new probability distribution $\vec{r}=\vec{p}\otimes\vec{q}\in\mathcal{P}_{nm}$ with marginal probabilities $\vec{p}$ and $\vec{q}$. Assuming that $\mathcal{I}=\{I_1,\cdots,I_L\}$ is a partition of $\{(1,1),\cdots,(1,m),\cdots,$ $(i,j),\cdots,(n,m)\}$, we have the aggregation $\vec{\epsilon}=\mathcal{I}(\vec{r})\in\mathcal{P}_L$ of $\vec{r}$ with $\epsilon_{k} = \sum_{(i,j)\in I_{k}}r_{(i,j)}=\sum_{(i,j)\in I_{k}}p_iq_j,k=1,\cdots,L$. In fact, the partitions $\mathcal{I}$, or equivalently the aggregations $\vec{\epsilon}$, that we are interested in are those that satisfy 
\begin{align}
\mathcal{I}(\vec{p}\otimes\vec{q})\prec\vec{p} \; , \; \mathcal{I}(\vec{p}\otimes\vec{q})\prec\vec{q} \; .
\label{eq:majorization_aggre}
\end{align}
Obviously, an identity belongs to this, i.e. $\mathcal{I}(\vec{p}\otimes\vec{q})=\vec{p}\otimes\vec{q} \prec \vec{p},\vec{q}$, because $p_i=\sum_{j}p_iq_j$ and $q_j=\sum_{i}p_iq_j$ are aggregations of $\vec{p}\otimes\vec{q}$. Note that \cref{eq:majorization_aggre} is a generalization of Lemma 1 in Ref. \cite{guhne04}. Specifically, every iteration of the degenerate case in Lemma 1 of Ref. \cite{guhne04} corresponds to a partition $\mathcal{I}$ with
\begin{align}
&I_{1} = \{(i,j),(k,l)|i\neq k \text{ and } j\neq l \} \; , \\ 
&I_{2},I_3\cdots,I_{nm-1} = \text{Identity} \; .
\end{align}
The non-steerable sates (by Alice) satisfy the mixed description of the joint probability for any two measurements $A_\mu,B_\mu$\cite{wiseman07}
\begin{align}
p(a_{\mu,i},b_{\mu,j}|A_{\mu},B_{\mu})_{\rho} = \sum_{\xi}\wp(a_{\mu,i}|A_{\mu},\xi)p(b_{\mu,j}|B_{\mu},\rho_{\xi})\wp_{\xi} \; .
\label{eq:lhs_steering}
\end{align}
Here, $\wp(a_i|A,\xi),\rho_{\xi}$ are the probability response function and local state of Alice and Bob, respectively; $\wp_{\xi}$ is a normalized distribution involving the hidden variable $\xi$ and $p(b_{\mu,j}|B_{\mu},\rho_{\xi})=\tr[\Pi_{j}^{B}\rho_{\xi}]$. For convenience, \cref{eq:lhs_steering} can be in form of
\begin{align}
\vec{p}(A_{\mu}\otimes B_{\mu})_{\rho} = \sum_{\xi}\vec{\wp}(A_{\mu},\xi)\otimes \vec{p}(B_{\mu},\rho_{\xi})\wp_{\xi} 
\end{align}
with $\vec{\wp}(A_{\mu},\xi)=(\wp(a_{\mu,i}|A_{\mu},\xi))$ and $\vec{p}(B_{\mu},\rho_{\xi})=(p(b_{\mu,j}|B_{\mu},\rho_{\xi}))$. If $\mathcal{I}$ is a partition of $\vec{p}\otimes\vec{q}$ satisfying $\mathcal{I}(\vec{p}\otimes\vec{q})\prec\vec{q}$, then we have for non-steerable states
\begin{align}
\mathcal{I}\left(\vec{p}(A_{\mu}\otimes B_{\mu})_{\rho}\right) = \sum_{\xi}\mathcal{I}\left(\vec{\wp}(A_{\mu},\xi)\otimes \vec{p}(B_{\mu},\rho_{\xi})\right)\wp_{\xi} \prec \sum_{\xi}\wp_{\xi}\vec{p}(B_{\mu},\rho_{\xi}) \; .
\end{align}
Since the binary operation $\odot\in\{\otimes,\oplus,+\}$ preserves the majorization relation, the majorization-based steering inequality is claimed
\begin{align}
\bigodot_{\mu}\mathcal{I}\left(\vec{p}(A_{\mu}\otimes B_{\mu})_{\rho}\right) \prec \sum_{\xi}\wp_{\xi}\bigodot_{\mu}\vec{p}(B_{\mu},\rho_{\xi}) \prec \vec{\omega}(\mathcal{B}) \; .
\label{eq:maj_steering_ineq}
\end{align}
Here, $\vec{\omega}(\mathcal{B})$ is the majorization UR bound of measurement bases $\mathcal{B}=(B_1,\cdots,B_N)$. In light of the local hidden state model of non-steerable states, Bob's measurements satisfy the majorization UR, while Alice's do not have any constraints. Thus, this result can be optimized by leveraging the symmetries inherent in the majorization UR bound and we have
\begin{align}
\bigodot_{\mu=1}^N\mathcal{I}\left(\vec{p}(\mathcal{J}(A_\mu)\otimes \mathcal{E}(B_\mu))_{\rho}\right) \prec \vec{\omega}(\mathcal{B}) \; .
\label{eq:steering_maj_ur_opt}
\end{align}
Here, $\mathcal{E}$ denotes the local transformations preserving the majorization UR bound $\vec{\omega}(\mathcal{B})$, i.e. $\vec{\omega}(\mathcal{E}(\mathcal{B}))=\vec{\omega}(\mathcal{B})$; $\mathcal{J}$ denotes any local transformation of Alice's measurements.

\section{Upper bounds of $\Omega_{L}$}\label{appen:Omega_N_approx_bound}

Here, we present some upper bounds of $\Omega_{L}$ for $N$ measurement bases. Given $N$ orthonormal and complete bases for $\mu=1,2,\cdots,N$, we define the index set $I=\{1,2,\cdots,d\}$ and the subsets $I_1,I_2,\cdots,I_N\subset I$, where $I_\mu$ denotes the index set of the $\mu$-th basis $A_\mu$. 
Assuming we choose $L$ base vectors from the $N$ bases, with each basis selecting $|I_\mu|$ vectors, we can define $L$ projectors $\{\ket{x_j}\bra{x_j}\}$, $j=1,2,\cdots,L=\sum_{\mu=1}^{N}|I_\mu|$ as follows:
\begin{align}
&\ket{x_{i_1}} = \ket{a_i^{(1)}} \; , \; \exists i\in I_1 \; , \; i_1=1,\cdots,|I_1| \; , \\ 
&\ket{x_{i_2}} = \ket{a_{i}^{(2)}} \; , \; \exists i\in I_2 \; , \; i_2=|I_1|+1,\cdots,|I_1|+|I_2| \; , \\ 
&\hspace{6ex}\vdots \notag \\
&\ket{x_{i_N}} = \ket{a_{i}^{(N)}} \; , \; \exists i\in I_N \; , \; i_N=\sum_{\mu=1}^{N-1}|I_\mu|+1,\cdots,L=\sum_{\mu=1}^{N}|I_\mu| \; .
\end{align}
With the help of these notations, a simple majorization UR bound yields \cite{friedland13,li19}
\begin{align}
\bigoplus_{\mu=1}^{N}\vec{p}(A_{\mu})       
\prec \vec{\omega} \; .
\end{align}
Here, $\vec{\omega}=(\Omega_{1},\Omega_2-\Omega_1,\cdots,\Omega_{Nd}-\Omega_{N(d-1)},0,\cdots,0)$ and $\Omega_{L}$ for $N$ measurements is defined as
\begin{align}
\Omega_{L}
&\equiv\max_{\substack{I_1,\cdots,I_N \\ \sum_{\mu=1}^{N}|I_\mu|=L}}\max_{\rho}\tr\left[\rho\left(\sum_{j=1}^L\ket{x_j}\bra{x_j}\right)\right] \\ 
&= \max_{\substack{I_1,\cdots,I_N \\ \sum_{\mu=1}^{N}|I_\mu|=L}}\left\|\sum_{j=1}^L\ket{x_j}\bra{x_j}\right\| \; .
\end{align}
Here, $\|\cdot\|$ denotes the operator norm, i.e. the largest singular value of the operator. Especially, if $L>d(N-1)$ we have $\Omega_L=N$ (the maximal value possible), which occurs if $I_1=\cdots=I_{N-1}=I$ and $I_N=\{l\}$, $\rho=\ket{a_l^{(N)}}\bra{a_l^{(N)}}$. The solution of $\Omega_L$ is a typical combinatorial optimization problem (COP), which has a discrete set of feasible solutions with a size of $\binom{Nd}{L}$. Next, we provides a few equivalent expressions of $\Omega_L$ for $N$ measurements.

\subsection*{Relaxation of optimization problem and the upper bound $\Gamma_L$}

$\Omega_L$ is defined as the following optimization problem $\Omega_{L} = \max_{\substack{I_1,\cdots,I_N \\ \sum_{\mu=1}^{N}|I_\mu|=L}}\left\|\sum_{j=1}^L\ket{x_j}\bra{x_j}\right\|$. Let $X\equiv\sum_{j=1}^L\ket{x_j}\bra{x_j}$ and $\lambda_1\geq\cdots\geq\lambda_d\geq0$ be the eigenvalues of $X$. 
In order to find an upper bound of $\Omega_L$, we consider the following relaxation of the optimization problem of $\Omega_L$
\begin{align}
\max \quad & \lambda_1 \notag \\
\textrm{s.t.} \quad & \sum_{i=1}^d\lambda_i=\tr[X]=L \; , \notag \\
& \sum_{i=1}^d\lambda_i^2=\tr[X^2]  \; , \label{eq:relaxation_omega} \\ 
&\quad \vdots \notag 
\end{align}
where, $\tr[X^2]=L+\sum_{i\neq j}|\braket{x_i|x_j}|^2=L+\sum_{\mu\neq\nu}\sum_{i\in I_\mu\\ j\in I_\nu}\left|\braket{a_i^{(\mu)}|a_j^{(\nu)}}\right|^2$. Obviously, the solution of the optimization problem \cref{eq:relaxation_omega} offers an upper bound of $\Omega_L$. In particular, if we only consider the first two constraints, i.e. $\sum_{i=1}^d\lambda_i=\tr[X]=L$ and $\sum_{i=1}^d\lambda_i^2=\tr[X^2]$, then we have the following upper bound
\begin{align}
\Omega_{L} \leq \Gamma_L 
\end{align}
with definition 
\begin{align}
\Gamma_L \equiv \max_{\substack{I_1,\cdots,I_N \\ \sum_{\mu=1}^{N}|I_\mu|=L}}\frac{L+\sqrt{(d-1)(d\tr[X^2]-L^2)}}{d} \; .
\end{align}

\begin{proof}
If we restrict our attention to the first two constraints, $\sum_{i=1}^d\lambda_i=\tr[X]=L$ and $\sum_{i=1}^d\lambda_i^2=\tr[X^2]$, we obtain the following optimization problem:
\begin{align}
\max \quad & \lambda_1 \notag \\
\text{s.t.} \quad & \sum_{i=1}^d\lambda_i=L, \\
& \sum_{i=1}^d\lambda_i^2=\tr[X^2], \\
& \lambda_1\geq\cdots\geq\lambda_d\geq0. \notag
\end{align}
To maximize $\lambda_1$, the remaining variables $\lambda_2, \cdots, \lambda_d$ must be as uniform as possible to minimize their contribution to the sum of squares. This implies setting $\lambda_2=\cdots=\lambda_d=b$ and $\lambda_1=a$ (with $a\geq b\geq0$). Substituting these into the constraints yields the quadratic equation
\begin{align}
da^2 -2La + L^2 - (d-1)\tr[X^2] = 0.
\end{align}
Solving for $a$ and taking the larger root, we find the maximum value $\lambda_1^* = \frac{L+\sqrt{(d-1)(d\tr[X^2]-L^2)}}{d}$.

We can rigorously verify this upper bound by contradiction. Suppose there exists a feasible solution with $\lambda_1' > \lambda_1^*$. From the first constraint, the sum of the remaining terms is $\sum_{i=2}^d\lambda_i^{\prime}=L-\lambda_1^{\prime}$. By the Cauchy-Schwarz inequality, the sum of their squares satisfies
\begin{equation}
\sum_{i=2}^d\lambda_i'^2 \geq \frac{1}{d-1}\left(\sum_{i=2}^d\lambda_i'\right)^2 = \frac{(L-\lambda_1')^2}{d-1}.
\end{equation}
Consequently, the total sum of squares for this hypothetical solution would be
\begin{equation}
\sum_{i=1}^d\lambda_i'^2 \geq \lambda_1'^2 + \frac{(L-\lambda_1')^2}{d-1}.
\end{equation}
Let $f(x) = x^2 + \frac{(L-x)^2}{d-1}$. This function is strictly increasing for $x \geq L/d$. Since $\lambda_1' > \lambda_1^* \geq L/d$ and $f(\lambda_1^*) = \tr[X^2]$, it follows that $f(\lambda_1') > \tr[X^2]$. This implies $\sum_{i=1}^d\lambda_i'^2 > \tr[X^2]$, which contradicts the second constraint. Thus, $\lambda_1$ cannot exceed $\lambda_1^*$. This completes the proof.
\end{proof}

\subsection*{Upper bounds based on the operator norm inequalities}

We first introduce the following lemma \cite{kittaneh97,schaffner07,tomamichel13,skrzypczyk15}:
\begin{lemma}\label{lem:ineq_operator_norm}
If $\{\ket{\psi_i}\}_{i=1}^n$ are $n$ normalized state vectors acting on an arbitrary $d$-dimensional Hilbert space $\mathcal{H}$, then we have the following inequalities 
\begin{align}
&\left\|\sum_{i=1}^n\ket{\psi_i}\bra{\psi_i}\right\| \leq \sqrt{\sum_{i,j=1}^{n}\left|\braket{\psi_i|\psi_j}\right|^2} \; , \label{eq:ineq1} \\ 
&\left\|\sum_{i=1}^n\ket{\psi_i}\bra{\psi_i}\right\| \leq 1 + (n-1)\max_{i\neq j}\left|\braket{\psi_i|\psi_j}\right| \; .
\label{eq:ineq2}
\end{align}
\end{lemma}
Let us define the following two quantities
\begin{align}
\Theta_L &\equiv 1 + (L-1)\max_{\mu\neq\nu}\left|\braket{a_i^{(\mu)}|a_j^{(\nu)}}\right| \; , \label{eq:theta} \\
\Lambda_L &\equiv \sqrt{L+\max_{\substack{I_1,\cdots,I_N \\ \sum_{\mu=1}^{N}|I_\mu|=L}}\sum_{\mu\neq\nu}\sum_{i\in I_\mu, j\in I_\nu}\left|\braket{a_i^{(\mu)}|a_j^{(\nu)}}\right|^2} \; , \label{eq:lambda} 
\end{align}
then we have $\Omega_L\leq \Theta_L$ and $\Omega_L\leq \Lambda_L$.
\begin{proof}
Employing \cref{eq:ineq1} in \cref{lem:ineq_operator_norm}, we have
\begin{align}
\Omega_{L} &= \max_{\substack{I_1,\cdots,I_N \\ \sum_{\mu=1}^{N}|I_\mu|=L}}\left\|\sum_{j=1}^L\ket{x_j}\bra{x_j}\right\| \\
&\leq \max_{\substack{I_1,\cdots,I_N \\ \sum_{\mu=1}^{N}|I_\mu|=L}}\sqrt{\sum_{i,j=1}^{L}\left|\braket{x_i|x_j}\right|^2} \; , \\
&= \sqrt{L+\max_{\substack{I_1,\cdots,I_N \\ \sum_{\mu=1}^{N}|I_\mu|=L}}\sum_{\mu\neq\nu}\sum_{\substack{i\in I_\mu\\ j\in I_\nu}}\left|\braket{a_i^{(\mu)}|a_j^{(\nu)}}\right|^2} \; .
\end{align}
In the final step, we utilized the orthonormality relation $\braket{a_i^{(\mu)}|a_j^{(\mu)}}=\delta_{ij}$ and the identity $\sum_{\mu=1}^N|I_\mu|=L$. $\Omega_L\leq \Theta_L$ is a straightforward consequence of \cref{eq:ineq2}.
\end{proof}

\subsection*{Upper bounds $\Gamma_L$, $\Lambda_L$ and $\Theta_L$ of $\Omega_L$ for $N$ MUBs.}
Assuming that $\{\ket{a_i^{(\mu)}}\}_{\mu=1}^N$ are $N$ $d$-dimensional MUBs, then we have 
\begin{align}
\left|\braket{a_i^{(\mu)}|a_j^{(\nu)}}\right|= \begin{cases}
1 / \sqrt{d} & , \; \mu \neq \nu \; ,\\ 
\delta_{i j} & , \;\mu=\nu \; .
\end{cases}
\label{eq:overlap_mub}
\end{align}
Substituting \cref{eq:overlap_mub} into $\tr[X^2]$, we have
\begin{align}
\max_{\substack{I_1,\cdots,I_N \\ \sum_{\mu=1}^{N}|I_\mu|=L}}\tr[X^2] &= L + \max_{\substack{I_1,\cdots,I_N \\ \sum_{\mu=1}^{N}|I_\mu|=L}}\sum_{\mu\neq\nu}\sum_{i\in I_\mu, j\in I_\nu}\left|\braket{a_i^{(\mu)}|a_j^{(\nu)}}\right|^2 \; , \\
&= L + \max_{\substack{I_1,\cdots,I_N \\ \sum_{\mu=1}^{N}|I_\mu|=L}}\sum_{\mu\neq\nu}\frac{|I_\mu||I_\nu|}{d} \; , \\
&= L+\max_{\substack{|I_1|,\cdots,|I_N| \\ \sum_{\mu=1}^{N}|I_\mu|=L}}\left(\sum_{\mu,\nu=1}^N\frac{|I_\mu||I_\nu|}{d}-\sum_{\mu=1}^N\frac{|I_\mu|^2}{d}\right) \; ,\\
&= L+\max_{\substack{|I_1|,\cdots,|I_N| \\ \sum_{\mu=1}^{N}|I_\mu|=L}}\left(\frac{L^2}{d}-\sum_{\mu=1}^N\frac{|I_\mu|^2}{d}\right) \; , \\
&= L+\frac{L^2}{d}-\min_{\substack{|I_1|,\cdots,|I_N| \\ \sum_{\mu=1}^{N}|I_\mu|=L}}\sum_{\mu=1}^N\frac{|I_\mu|^2}{d} \; , \\
&= L + \frac{L^2-\Phi(L)}{d} \; , 1\leq L \leq N(d-1) \; ,
\end{align}
where $\Phi(L)=N\lfloor\frac{L}{N}\rfloor^2+\left(L-N\lfloor\frac{L}{N}\rfloor\right)\left(2\lfloor\frac{L}{N}\rfloor+1\right)$ with the floor function $\lfloor\cdot\rfloor$. Substituting this into $\Gamma_L$, we have
\begin{align}
\Gamma_L &= \max_{\substack{I_1,\cdots,I_N \\ \sum_{\mu=1}^{N}|I_\mu|=L}}\frac{L+\sqrt{(d-1)(d\tr[X^2]-L^2)}}{d} \; , \\ 
&= \frac{L+\sqrt{(d-1)(dL-\Phi(L))}}{d} \; , \; 1\leq L\leq N(d-1) \; .
\end{align}
Substituting \cref{eq:overlap_mub} into $\Lambda_L$ and $\Theta_L$ respectively, we have
\begin{align}
\Lambda_L &= \sqrt{L+\max_{\substack{I_1,\cdots,I_N \\ \sum_{\mu=1}^{N}|I_\mu|=L}}\sum_{\mu\neq\nu}\sum_{\substack{i\in I_\mu\\ j\in I_\nu}}\left|\braket{a_i^{(\mu)}|a_j^{(\nu)}}\right|^2} \; , \\
&= \sqrt{\frac{dL+L^2-\Phi(L)}{d}} \; , \; 1\leq L\leq N(d-1) \; .
\end{align}
Finally, we formulate three analytical upper bounds of $\Omega_L$ for $N$ MUBs as follows
\begin{align}
\begin{cases}
&\Gamma_L = \frac{L+\sqrt{(d-1)(dL-\Phi(L))}}{d} \; , \; 1\leq L\leq N(d-1)  \; , \\
&\Lambda_L = \sqrt{\frac{dL+L^2-\Phi(L)}{d}} \; , \; 1\leq L\leq N(d-1) \; , \\ 
&\Theta_L = 1 + \frac{L-1}{\sqrt{d}} \; , \; 1\leq L\leq N(d-1) \; .
\end{cases}
\end{align}
It is worth noting that though $\Theta_L$ is tighter than $\Lambda_L$ for MUBs scenario, $\Lambda_L$ provides a tighter upper bound than $\Theta_L$ for general $N$ measurement bases. For a complete set of $d+1$ MUBs, $\bar{\Gamma}_N$ provides a tighter upper bound than $\bar{\Lambda}_N$ and $\bar{\Theta}_N$ as shown in \supptabref{tab:upper_bound_mub}.
\begin{table}
\caption{\label{tab:upper_bound_mub}The comparison of the upper bounds $\bar{\Omega}_N$, $\bar{\Theta}_N$, $\bar{\Gamma}_N$ and $\bar{\Lambda}_N$ for $N$ MUBs. Blue values indicate the optimal $\bar{\Omega}_N$.}
\centering
\newcommand{\fv}[4]{\textcolor{blue}{#1}/\textcolor{red}{#2}/\textcolor{ForestGreen}{#3}/\textcolor{purple}{#4}}
\renewcommand{\arraystretch}{1.4}
\setlength{\tabcolsep}{3pt} 

\resizebox{\textwidth}{!}{%
\begin{tabular}{c cccccc}
\toprule
& \multicolumn{6}{c}{\fv{$\bar{\Omega}_N$}{$\bar{\Theta}_N$}{$\bar{\Gamma}_N$}{$\bar{\Lambda}_N$}} \\
\cmidrule(l){2-7}
$N$ & $d=2$ & $d=3$ & $d=4$ & $d=5$ & $d=6$ & $d=7$ \\
\midrule
2 & \fv{0.8536}{0.8536}{0.8536}{0.8660} & \fv{0.7887}{0.7887}{0.8047}{0.8165} & \fv{0.7500}{0.7500}{0.7803}{0.7906} & \fv{0.7236}{0.7236}{0.7657}{0.7746} & \fv{0.7041}{0.7041}{0.7559}{0.7638} & \fv{0.6890}{0.6890}{0.7489}{0.7560} \\
3 & \fv{0.7887}{0.8047}{0.7887}{0.8165} & \fv{0.7124}{0.7182}{0.7182}{0.7454} & \fv{0.6667}{0.6667}{0.6830}{0.7071} & \fv{0.6315}{0.6315}{0.6619}{0.6831} & \fv{0.6030}{0.6055}{0.6478}{0.6667} & \fv{0.5846}{0.5853}{0.6377}{0.6547} \\
4 & -- & \fv{0.6545}{0.6830}{0.6667}{0.7071} & \fv{0.6250}{0.6250}{0.6250}{0.6614} & \fv{0.5693}{0.5854}{0.6000}{0.6325} & -- & \fv{0.5276}{0.5335}{0.5714}{0.5976} \\
5 & -- & -- & \fv{0.5732}{0.6000}{0.5854}{0.6325} & \fv{0.5343}{0.5578}{0.5578}{0.6000} & -- & \fv{0.4960}{0.5024}{0.5262}{0.5606} \\
6 & -- & -- & -- & \fv{0.5090}{0.5393}{0.5266}{0.5774} & -- & \fv{0.4743}{0.4816}{0.4928}{0.5345} \\
7 & -- & -- & -- & -- & -- & \fv{0.4587}{0.4668}{0.4668}{0.5151} \\
8 & -- & -- & -- & -- & -- & \fv{0.4272}{0.4557}{0.4460}{0.5000} \\
\bottomrule
\end{tabular}%
}
\end{table}

\section{Majorization steering inequalities in some scenarios}\label{appen:steering_inequalities}
Here, we present the majorization steering inequalities for 2-qubit states, isotropic states and Werner states. We consider the partition $\mathcal{I}=\{I_1,\cdots,I_d\}$ of $\{(1,1),\cdots,(1,n),\cdots,$ $(i,j),\cdots,(d,d)\}$ with elements
\begin{align}
I_k = \{(i,j) | j\equiv(i+k-1)\pmod d,1\leq i,j \leq d\} \; ,
\label{eq:partition}
\end{align}
where $k=1,\cdots,d$. The partition $\mathcal{I}$ results in an aggregation $\vec{\Upsilon}$ of $p(a_i,b_j|A,B;\rho)$ with 
\begin{align}
\Upsilon_k(A,B;\rho) = \sum_{(i,j)\in I_k} p(a_i,b_j|A,B;\rho) = \frac{1}{d} + \frac{1}{4}\sum_{(i,j)\in I_k}(\vec{a}_i,\mathcal{T}\vec{b}_j) \; ,
\end{align}
where $\vec{a}_i\left(\vec{b}_j\right)$ is Bloch vector of projectors and $\mathcal{T}_{\mu\nu}=\tr[\rho \left(\pi_{\mu}\otimes\pi_{\nu}\right)]$ is correlation matrix of $\rho$. Given observables $\mathcal{A}=(A_1,\cdots,A_N)$ and $\mathcal{B}=(B_1,\cdots,B_N)$, Theorem 1 in the main text yields the following steering inequality:
\begin{align}
\bigoplus_{\mu=1}^{N}\vec{\Upsilon}(\mathcal{J}(A_\mu),\mathcal{E}(B_{\mu});\rho) \prec\vec{\omega}(\mathcal{B}) \; ,
\end{align}
where the joint probability is 
\begin{align}
\Upsilon_{k}(\mathcal{J}(A_\mu),\mathcal{E}(B_{\mu});\rho)&= \sum_{(i,j)\in I_k} p(a_{\mu,i},b_{\mu,j}|\mathcal{J}(A_\mu),\mathcal{E}(B_{\mu});\rho) \\ 
&= \frac{1}{2} + \frac{1}{4}\sum_{(i,j)\in I_k}(\vec{a}_{\mu,i}^{\,\prime},\mathcal{T}\vec{b}_{\mu,j}^{\prime}) \; .
\end{align}
Here, $\vec{a}_{\mu,i}^{\,\prime}$ ($\vec{b}_{\mu,j}^{\prime}$) denotes Bloch vector of projector of $\mathcal{J}(A_\mu)$ ($\mathcal{E}(B_\mu)$). Immediately, a based-aggregation $\vec{\Upsilon}$ steering inequality yields
\begin{align}
\mathcal{S}_L \equiv \frac{1}{L}\sum_{k=1}^{L}\left[\bigoplus_{\mu}^{N}\vec{\Upsilon}_{\mu}\right]_{k}^{\downarrow} \leq \bar{\Omega}_{L} \; , \; L=2,\cdots,Nd \; .
\end{align}
Here $[\cdot]_{k}^{\downarrow}$ denotes the $k$-th largest component of a vector and $\bar{\Omega}_{L}=\frac{\Omega_{L}}{L}$ with $\Omega_{L}=\sum_{k=1}^{L}\left[\vec{\omega}(\mathcal{B})\right]_{k}^{\downarrow}$. We have defined the quantity $\mathcal{S}_L$ as the steering parameter for $N$ measurement settings in arbitrary dimensional Hilbert space.

\subsection*{Two qubits system}

For qubit dichotomy measurements $A_{\mu}=\vec{a}_{\mu}\cdot\vec{\sigma}$ and $B_{\mu}=\vec{b}_{\mu}\cdot\vec{\sigma}$ with $\mu=1,\cdots,N$, we have $i,j=\pm$ and $\vec{a}_{\mu,+}=-\vec{a}_{\mu,-}=\vec{a}_{\mu}$ and similarly for $\vec{b}_{\mu,j}$, which results in 
\begin{align}
\Upsilon_{k}(\mathcal{J}(A_\mu),\mathcal{E}(B_{\mu});\rho)= \frac{1}{2}\left[1\pm(\vec{a}_{\mu}^{\,\prime},\mathcal{T}\vec{b}_{\mu}^{\prime})\right] \; , \; k=\pm \; ,
\end{align}
and $\Omega_{N}=\frac{N}{2} + \frac{1}{2}\max_{i_{\mu}\in\{\pm\}}\left|\sum_{\mu=1}^{N}\vec{b}_{\mu,i_\mu}\right|$ with $\vec{b}_{\mu,+}=-\vec{b}_{\mu,-}=\vec{b}_{\mu}$.

Now, we consider the local transformation $\mathcal{E}(B_\mu)=\vec{b}_{\mu}^{\prime}\cdot\vec{\sigma}$ with $\vec{b}_{\mu}^{\prime}=O_B\vec{b}_{\mu}$ and $O_B\in O(3)$, which preserves the majorization UR bound, obviously. Simultaneously, we set the local transformation $\mathcal{J}(A_\mu)=\vec{a}_{\mu}^{\,\prime}\cdot\vec{\sigma}$ with $\vec{a}_{\mu}^{\,\prime}=O_A\vec{a}_{\mu}$ and $O_A\in O(3)$, whereupon we have
\begin{align}
\Upsilon_{\pm}(\mathcal{J}(A_\mu),\mathcal{E}(B_{\mu});\rho)&= \frac{1}{2}\left[1\pm(O_A\vec{a}_{\mu},\mathcal{T}(O_B\vec{b}_{\mu}))\right]  \\ 
&= \frac{1}{2}\left[1\pm(\vec{a}_{\mu},(O_A^{\mathrm{T}}\mathcal{T}O_B)\vec{b}_{\mu})\right] \; , \; O_A,O_B\in O(3) \; .
\end{align}
Let the local transformations $\mathcal{J}$ and $\mathcal{E}$ conduct the singular value decomposition of $\mathcal{T}$, i.e. $O_A^{\mathrm{T}}\mathcal{T}O_B=\operatorname{diag}\{t_1,t_2,t_3\}$. Then we have
\begin{align}
\Upsilon_{\pm}(\mathcal{J}(A_\mu),\mathcal{E}(B_{\mu});\rho)= \frac{1}{2}\left[1 \pm (\vec{t}\circ\vec{a}_\mu)\cdot\vec{b}_{\mu}\right] \; .
\end{align}
Here, $\vec{t}=(t_1,t_2,t_3)$ is singular value vector of correlation matrix $\mathcal{T}$ of quantum state $\rho$ and $\circ$ denotes Hadamard product. 

Considering that 2-qubit Werner state $\rho_{\mathrm{W}}=(1-w)\mathds{1}/4+w\ket{\psi_{-}}\bra{\psi_{-}}$ with $\ket{\psi_{-}} = \frac{1}{\sqrt{2}}(\ket{01} - \ket{10})$ and setting $\vec{a}_{\mu}=\vec{b}_{\mu}$, then we have $\Upsilon_{\pm}=\frac{1\pm w}{2}$ and the violation of the following inequality signifies the steerability of 2-qubit Werner states:
\begin{align}
w \leq 2\bar{\Omega}_{L}-1 \; , \; L=2,\cdots,N \; . 
\label{eq:steering_2-qubit_werner}
\end{align}
We note that Saunders \emph{et al.} formulated the following linear steering inequality \cite{saunders10}
\begin{align}
\frac{1}{N}\sum_{k=1}^N|\braket{A_k\otimes B_k}| \leq C_N \; ,
\label{eq:saunders_steering}
\end{align}
where $C_N=\frac{1}{N}\max_{\{a_k\}}\left[\lambda_{\max}\left(\sum_{k=1}^{N}a_kB_k\right)\right]$ and $a_k=\pm1$. $\lambda_{\max}(X)$ denotes the largest eigenvalue of $X$. It is evident that our approach, along with Saunders \emph{et al.}'s inequality \cref{eq:saunders_steering}, is equivalent for 2-qubit Werner states due to $\bar{\Omega}_N=\frac{1}{2}+C_N$ (readily verified by employing the Bloch representation of $\Omega_N$).

\begin{figure}
\centering
\includegraphics[width=0.5\textwidth]{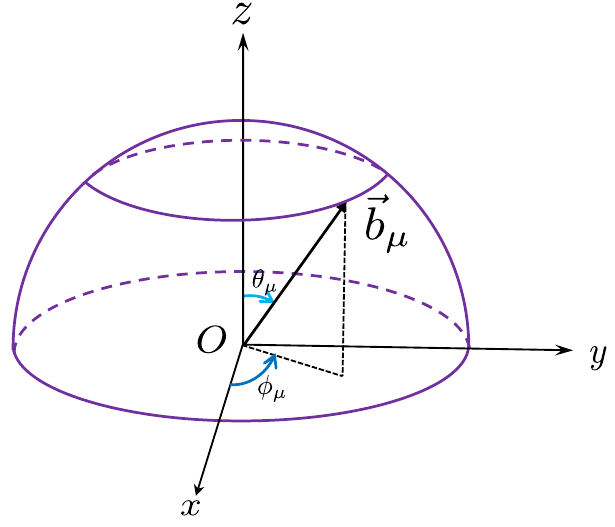}
\caption{Infinite measurement settings. Measurements are performed for the whole hemisphere i.e., the all $\vec{b}_\mu$ with $\forall \theta_\mu\in[0,\pi/2],\phi_\mu\in[0,2\pi]$. }
\label{fig:inf_measurement_settings}
\end{figure}

\emph{Infinite measurement settings scenario.} \cref{eq:steering_2-qubit_werner} offers an elegant method to discuss the steerability involving with the infinite measurement settings, which corresponds to calculate the limitation $\lim_{N\to\infty}\frac{\Omega_N}{N}$ when measurements are performed for the entire hemisphere as depicted in \suppfigref{fig:inf_measurement_settings}. In this case, the limit $\lim_{N\to\infty}\frac{\Omega_N}{N}$ can be calculated as follows:
\begin{align}
\lim_{N\to\infty}\frac{\Omega_N}{N} &= \frac{1}{2}+\frac{1}{2}\lim_{N\to\infty}\frac{|\sum_{\mu=1}^N\vec{b}_\mu|}{N} \\
&= \frac{1}{2}+ \frac{1}{2}\lim_{N\to\infty}\sqrt{L_x^2+L_y^2+L_z^2} \; ,
\end{align}
where we have set $\vec{L}=\frac{\sum_{\mu=1}^N\vec{b}_{\mu}}{N}$. Using spherical coordinates it is easy to see that $\lim_{N\to\infty}L_x=\frac{1}{2\pi}\lim_{\Delta\Omega\to 0}\sum_{\mu=1}^{2\pi/\Delta\Omega}\sin^2\theta_\mu\cos\phi_\mu\Delta\theta_{\mu}\Delta\phi_\mu=\frac{1}{2\pi}\int_0^{\pi/2}\int_0^{2\pi}\sin^2\theta\cos\phi d\theta d\phi=0$ with $\Delta\Omega=\sin\theta\Delta\theta\Delta\phi$ and $N=\frac{2\pi}{\Delta\Omega}$. Similarly, we obtain $\lim_{N\to\infty}L_y=0$ and $\lim_{N\to\infty}L_z=\frac{1}{2}$. Thus, we have $\lim_{N\to\infty}\frac{\Omega_N}{N}=\frac{3}{4}$, which achieves the critical value $w=\frac{1}{2}$ of steerability for the two-qubit Werner states. Especially, when the measurements are constrained to the semi-circle, i.e. $\phi_\mu\in[0,\pi/2],\theta_{\mu}=\frac{\pi}{2}$, we have $\lim_{N\to\infty}\frac{\Omega_N}{N}=\frac{\pi+2}{2\pi}$ and the threshold $w=\frac{2}{\pi}$ of steerability.

\subsection*{The isotropic states and Werner states}

\emph{Bloch representation and description of quantum states.} We now focus on the arbitrary dimensional Werner and isotropic states. The isotropic states are defined as \cite{horodecki99}
\begin{align}
\rho_{\mathrm{ISO}} = \frac{1-F}{d^2-1}\mathds{1}\otimes\mathds{1} + \frac{d^2F-1}{d^2-1}P_{+} \; .
\label{eq:isotropic_state}
\end{align}
Here, $F\in[0,1]$ is the fidelity of the state $\rho_{\mathrm{ISO}}$ with respect to the maximally entangled state $\ket{\psi_{+}}=\frac{1}{\sqrt{d}}\sum_{m=1}^d\ket{mm}$, i.e. $F=\tr[\rho_{\mathrm{ISO}}P_{+}]$ with $P_{+}=\ket{\psi_{+}}\bra{\psi_{+}}$. An alternative form of the isotropic states is given by the maximally entangled states with white noise, i.e. $\rho_{\mathrm{ISO}}=\frac{1-w}{d^2}\mathds{1}\otimes\mathds{1}+w\ket{\psi_{+}}\bra{\psi_{+}}$ with $w=\frac{d^2F-1}{d^2-1}$. The Werner states are defined as \cite{wiseman07}
\begin{align}
\rho_{\mathrm{W}} = \left(\frac{d-1+\eta}{d-1}\right)\frac{\mathds{1}\otimes \mathds{1}}{d^2} - \left(\frac{\eta}{d-1}\right)\frac{\mathds{V}}{d} \; ,
\label{eq:werner_state}
\end{align}
where $\mathds{V}$ is the swap operator defined as $\mathds{V}\ket{\psi}\otimes \ket{\varphi} := \ket{\varphi}\otimes \ket{\psi},\forall \ket{\psi},\ket{\varphi}$ and $\eta\in[0,1]$. It is straightforward to convert \cref{eq:isotropic_state,eq:werner_state} into their Bloch representations
\begin{align}
\rho_{\mathrm{ISO}} &= \frac{1}{d^2} \mathds{1}\otimes \mathds{1} + \frac{1}{4} \sum_{\mu=1}^{d^2-1} \frac{2w}{d} \pi_{\mu} \otimes \pi_{\mu}^{\mathrm{T}} \; , \\ 
\rho_{\mathrm{W}} &= \frac{1}{d^2} \mathds{1} \otimes \mathds{1} + \frac{1}{4}\sum_{\mu=1}^{d^2-1} \frac{-2\eta}{d(d-1)} \pi_{\mu} \otimes \pi_{\mu}\; .
\end{align}
Without loss of generality, one can set the first $N_{\mathrm{anti}}=d(d-1)/2$ generators are to be antisymmetric and the last $N_{\mathrm{sym}}=d(d+1)/2-1$ generators are to be symmetric, i.e. $\pi_{\mu}^{\mathrm{T}}=-\pi_{\mu}$ for $\mu=1,\cdots,N_{\mathrm{anti}}$ and $\pi_{\mu}^{\mathrm{T}}=\pi_{\mu}$ for $\mu=N_{\mathrm{anti}}+1,\cdots,d^2-1$ \cite{pfeifer03}. Then the correlation matrices $\mathcal{T}$ of the isotropic states and Werner states are given by $\mathcal{T}_{\mathrm{ISO}}=\frac{2w\Phi}{d}$ with $\Phi=\big[\begin{smallmatrix}
-\mathds{1}_{N_{\mathrm{anti}}} & 0\\
0 & \mathds{1}_{N_{\mathrm{sym}}}
\end{smallmatrix}\big]$ and $\mathcal{T}_{\mathrm{W}}=\frac{-2\eta\mathds{1}}{d(d-1)}$, respectively.

\emph{Measurement settings.} Without loss of generality, one can set $A_{\mu}=\vec{a}_{\mu}\cdot\vec{\pi}$ and $B_{\mu}=\vec{b}_{\mu}\cdot\vec{\pi}$ with $\mu=1,\cdots,N$, where $\vec{a}_\mu,\vec{b}_\mu$ are $(d^2-1)$ dimensional real vectors. After the aggregation with partition \cref{eq:partition}, the joint probability reads as
\begin{align}
\Upsilon_{k}(\mathcal{J}(A_\mu),\mathcal{E}(B_{\mu});\rho)
= \frac{1}{d} + \frac{1}{4}\sum_{(i,j)\in I_k}(\vec{a}_{\mu,i}^{\,\prime},\mathcal{T}\vec{b}_{\mu,j}^{\prime}) \; .
\end{align}
The correlation matrix of the Werner state is a constant matrix and hence the optimal local transformations $\mathcal{J}$ and $\mathcal{E}$ should be the identity operation, which leads to the joint probability
\begin{align}
\Upsilon_{k}^{\mathrm{W}}(\mathcal{J}(A_\mu),\mathcal{E}(B_{\mu});\rho)
= \frac{1}{d} + \frac{-\eta}{2d(d-1)}\sum_{(i,j)\in I_k}(\vec{a}_{\mu,i},\vec{b}_{\mu,j}) \; .
\end{align}
Setting $A_\mu=B_\mu$ and employing $(\vec{b}_{\mu,i},\vec{b}_{\mu,j})=2(d\delta_{ij}-1)/d$ due to the orthonormality $\braket{b_{\mu,i}|b_{\mu,j}}=\delta_{ij}$, we have
\begin{align}
\Upsilon_{k}^{\mathrm{W}}(\mathcal{J}(A_\mu),\mathcal{E}(B_{\mu});\rho)
&= \frac{1}{d} + \frac{-\eta}{2d(d-1)}\sum_{(i,j)\in I_k}(\vec{b}_{\mu,i},\vec{b}_{\mu,j}) \\ 
&= \left\{
\begin{aligned}
&\frac{1-\eta}{d} \; , \; k=1 \; , \\ 
&\frac{d-1+\eta}{d(d-1)} \; , \; k=2,\cdots,d \; .
\end{aligned}
\right.
\end{align}
Next, we consider the isotropic states whose joint probability reads 
\begin{align}
\Upsilon_{k}^{\mathrm{ISO}}(\mathcal{J}(A_\mu),\mathcal{E}(B_{\mu});\rho)
= \frac{1}{d} + \frac{w}{2d}\sum_{(i,j)\in I_k}(\vec{a}_{\mu,i}^{\,\prime},\Phi\vec{b}_{\mu,j}^{\prime}) \; .
\end{align}
Setting $A_\mu=B_\mu$ and the local transformations $\mathcal{E}(B_\mu)=B_\mu,\mathcal{J}(A_\mu)=B_\mu^\prime=\vec{b}_{\mu}^{\,\prime}\cdot\vec{\pi}$ with $\vec{b}_{\mu}^{\,\prime}=\Phi\vec{b}_{\mu}$, we have
\begin{align}
\Upsilon_{k}^{\mathrm{ISO}}(\mathcal{J}(A_\mu),\mathcal{E}(B_{\mu});\rho)
&= \frac{1}{d} + \frac{w}{2d}\sum_{(i,j)\in I_k}(\vec{b}_{\mu,i},\Phi^2\vec{b}_{\mu,j}) \\ 
&= \frac{1}{d} + \frac{w}{2d}\sum_{(i,j)\in I_k}(\vec{b}_{\mu,i},\vec{b}_{\mu,j}) \\ 
&= \left\{
\begin{aligned}
&\frac{1+(d-1)w}{d} \; , \; k=1 \; , \\ 
&\frac{1-w}{d} \; , \; k=2,\cdots,d \; .
\end{aligned}
\right.
\end{align}
\emph{Majorization steering inequalities.} In light of the results above, we have
\begin{align}
\left\{
\begin{aligned}
&\mathcal{S}_L^{\mathrm{W}} = \frac{1}{L}\sum_{k=1}^{L}\left[\bigoplus_{\mu}^{N}\vec{\Upsilon}_{\mu}^{\mathrm{W}}\right]_{k}^{\downarrow} = \frac{d-1+\eta}{d(d-1)} \; , \; L=2,\cdots,N(d-1) \; , \\
&\mathcal{S}_L^{\mathrm{ISO}} = \frac{1}{L}\sum_{k=1}^{L}\left[\bigoplus_{\mu}^{N}\vec{\Upsilon}_{\mu}^{\mathrm{ISO}}\right]_{k}^{\downarrow} = \frac{1+(d-1)w}{d} \; , \; L=2,\cdots,N \; .
\end{aligned}
\right.
\end{align}
Therefore, the majorization steering inequalities for Werner and isotropic states are given by
\begin{align}
\left\{
\begin{aligned}
&w \leq \frac{1}{d-1}\left(d\bar{\Omega}_{L}-1\right) \; , \; L=2,\cdots,N \; , &\text{for isotropic states} \; , \\  
&\eta \leq (d-1)\left(d\bar{\Omega}_{L}-1\right) \; , \; L=2,\cdots,N(d-1) \; , &\text{for Werner states} \; .
\end{aligned}
\right.
\end{align}
Obviously, we recover the qubit result $w\leq 2\bar{\Omega}_{L}-1$ for $d=2$ as expected.


\end{document}